\documentclass[acmtog,anonymous=false,timestamp=false,review=false ]{acmart}
\acmSubmissionID{0246}

\usepackage{booktabs} 
\usepackage{bm}
\usepackage{comment}
\usepackage{enumitem}
\usepackage{lipsum}
\usepackage{overpic}
\usepackage{mathtools}
\usepackage{mdframed}

\setcitestyle{square}

\usepackage[utf8]{inputenc}

\usepackage{algorithm}
\usepackage{algpseudocode}
\algtext*{EndWhile}
\algtext*{EndIf}
\algtext*{EndProcedure}
\usepackage{multirow}
\usepackage{colortbl}
\usepackage{xcolor}

\usepackage{wrapfig}

\newlength\savedwidth
\newcommand\whline[1]{\noalign{\global\savedwidth\arrayrulewidth
                               \global\arrayrulewidth #1} %
                      \hline
                      \noalign{\global\arrayrulewidth\savedwidth}}

\setcopyright{acmlicensed}

\definecolor{blue}{HTML}{1F78E4}
\newcommand{\camera}[1]{#1}

\newcommand{\figref}[1]{Figure~\ref{fig:#1}}

\newcommand{\eqnref}[1]{Eq.~\eqref{eq:#1}}
\newcommand{\secref}[1]{\S\ref{sec:#1}}
\newcommand{\eq}[1]{\eqref{eq:#1}}

\renewcommand{\vec}[1]{\bm{#1}}

\newcommand{\txt}[1]{\textrm{#1}}

\newcommand{\mdots}{...}

\newcommand{\tsd}{360\textdegree{} }

\setcopyright{acmlicensed}
\acmJournal{TOG}
\acmYear{2018}\acmVolume{37}\acmNumber{4}\acmArticle{111}\acmMonth{8} \acmDOI{10.1145/3197517.3201391}

\begin{document}
\title{Scene-Aware Audio for 360\textdegree{} Videos}

\author{Dingzeyu Li}
\orcid{0000-0002-4222-8105}
\affiliation{
\institution{Columbia University}
}
\email{dli@cs.columbia.edu}

\author{Timothy R. Langlois}
\orcid{0000-0002-5043-8698} 
\affiliation{%
\institution{Adobe Research}
}
\email{tlangloi@adobe.com}

\author{Changxi Zheng}
\orcid{0000-0001-9228-1038}
\affiliation{
\institution{Columbia University}
}
\email{cxz@cs.columbia.edu}

\renewcommand\shortauthors{Li, D. et al}

\begin{abstract}
Although \tsd cameras ease the capture of panoramic footage, it remains
challenging to add realistic \tsd audio that blends into the captured scene and
is synchronized with the camera motion. We present a method for adding
scene-aware spatial audio to \tsd videos in typical indoor scenes, using only a
conventional mono-channel microphone and a speaker. We observe that the late
reverberation of a room's impulse response is usually diffuse spatially and
directionally. Exploiting this fact, we propose a method that synthesizes the
directional impulse response between any source and listening locations by
combining a synthesized early reverberation part and a measured late
reverberation tail.  The early reverberation is simulated using a geometric
acoustic simulation and then enhanced using a frequency modulation method to
capture room resonances. The late reverberation is extracted from a recorded
impulse response, with a carefully chosen time duration that separates out the
late reverberation from the early reverberation. In our validations, we show
that our synthesized spatial audio matches closely with recordings using
ambisonic microphones. Lastly, we demonstrate the strength of our method in
several applications.
\end{abstract}

%
%
%

%
%



\maketitle

\section{Introduction}

The ecosystem of \tsd video is flourishing. Devices such as
the Samsung Gear 360 and the Ricoh Theta
have facilitated \tsd video capture;
software such as Adobe Premiere Pro has included features for
editing \tsd panoramic footage;
and online platforms such as Youtube and Facebook have promoted easy sharing
and viewing of \tsd content.
With these technological advances, video creators now have a whole new set
of tools
for creating immersive visual experiences.
Yet, the creation of their auditory accompaniment, the immersive audio, is not as easy.
Immersive \tsd videos are noticeably lacking immersive scene-aware \tsd audio.

\begin{figure}[t]
    \includegraphics[width=0.99\columnwidth]{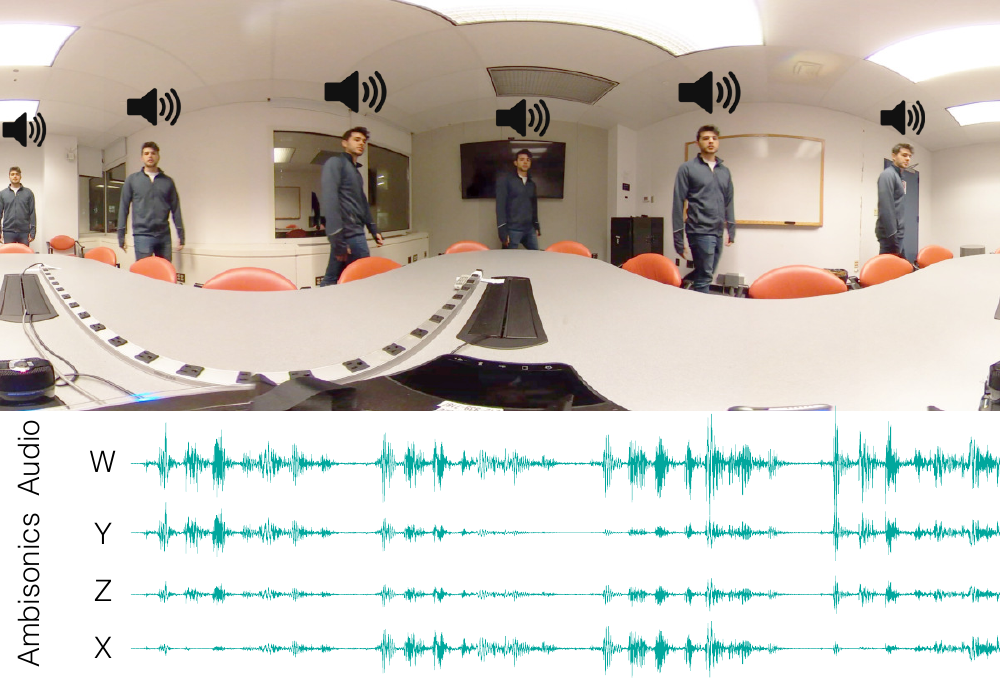}
    \vspace{-2mm}
    \caption{ \textbf{\tsd audiovisual capture.}
    Our method enables video creators to add ambisonic audio (bottom) in a \tsd video of
    a indoor scene (top).
    When viewers watch the video and change the camera angle, they hear the binaural
    audio consistent with the current viewing angle. Our method has no restriction on
    the input audio. In the case shown here, the input audio is the person's speech
    captured using a conventional mono-channel microphone collocated with the \tsd camera,
    and our method converts the mono-channel input audio into a spatial audio in standard
    ambisonic format. The waveforms show a first-order ambisonic output (four channels),
    although our method supports an arbitrary order of ambisonics.
    \label{fig:teaser}}
    \vspace{-3mm}
\end{figure}

Toward filling this gap, we propose a method that enables \tsd video creators to easily
add spatial audio from specified sound sources in a typical indoor scene, such
as the conference room shown in \figref{teaser}.
Our method consists of two stages.
We first record a single acoustic impulse response in a room using a readily available
mono-channel microphone and a simple setup.
Then, provided any \tsd footage captured in the same environment and
a piece of source audio, our method
outputs the \tsd video with an accompanying ambisonics spatial soundtrack.
The resulting soundfield captures the spatial sound effects at the camera location,
even if the camera is dynamic, as if the input audio is emitted from a user-specified sound source in
the environment.
Our method has no restriction on the input audio: it could be
artificially synthesized, recorded in an anechoic chamber,  or
recorded in the same scene simply using a conventional mono-channel microphone (\figref{teaser}).
\camera{The generated ambisonic audio can be directly played back in realtime
by a spatial audio player.}

A conventional microphone and a sound source are the only requirements our method, 
in addition to a \tsd camera.
This contrasts starkly with the current approach of
capturing spatial audio, which requires the use of a soundfield ambisonic microphone, which uses a microphone array consisting of multiple carefully positioned mono-channel microphones to record the spatial sound field.
These devices are generally expensive, and
currently very few \tsd cameras have an integrated ambisonic microphone.
When designing sound for traditional media,
audio from each source is processed to add various effects:
noise removal, frequency equalization, dynamic compression, panning, and so forth.
Then, the audio clips are mixed into a cohesive soundtrack for a specific layout
of speakers, with \emph{a fixed camera angle}. 
While ambisonics could be created
virtually from these sources, it is only feasible to do this manually in a space
with no reflections. In real rooms reflections off of surfaces and sources need to
account for direction. Our method provides an easy way to achieve these directed reflections.

Our method enables \tsd video creators
to incorporate spatial audio at a lower cost, without the need of ambisonic microphones.
More importantly, it allows the creator to reuse the well-established audio production
pipeline, where sound effects are designed, recorded, denoised, and composed ---
without worrying about downstream ambisonic effects.
Afterwards, our method automatically incorporates room acoustic effects in the video-shooting scene,
and converts the sound produced in the earlier stage into spatial audio, which is fully
synchronized with the camera trajectory in the \tsd video.

\begin{figure}[t]
    \includegraphics[width=0.90\columnwidth]{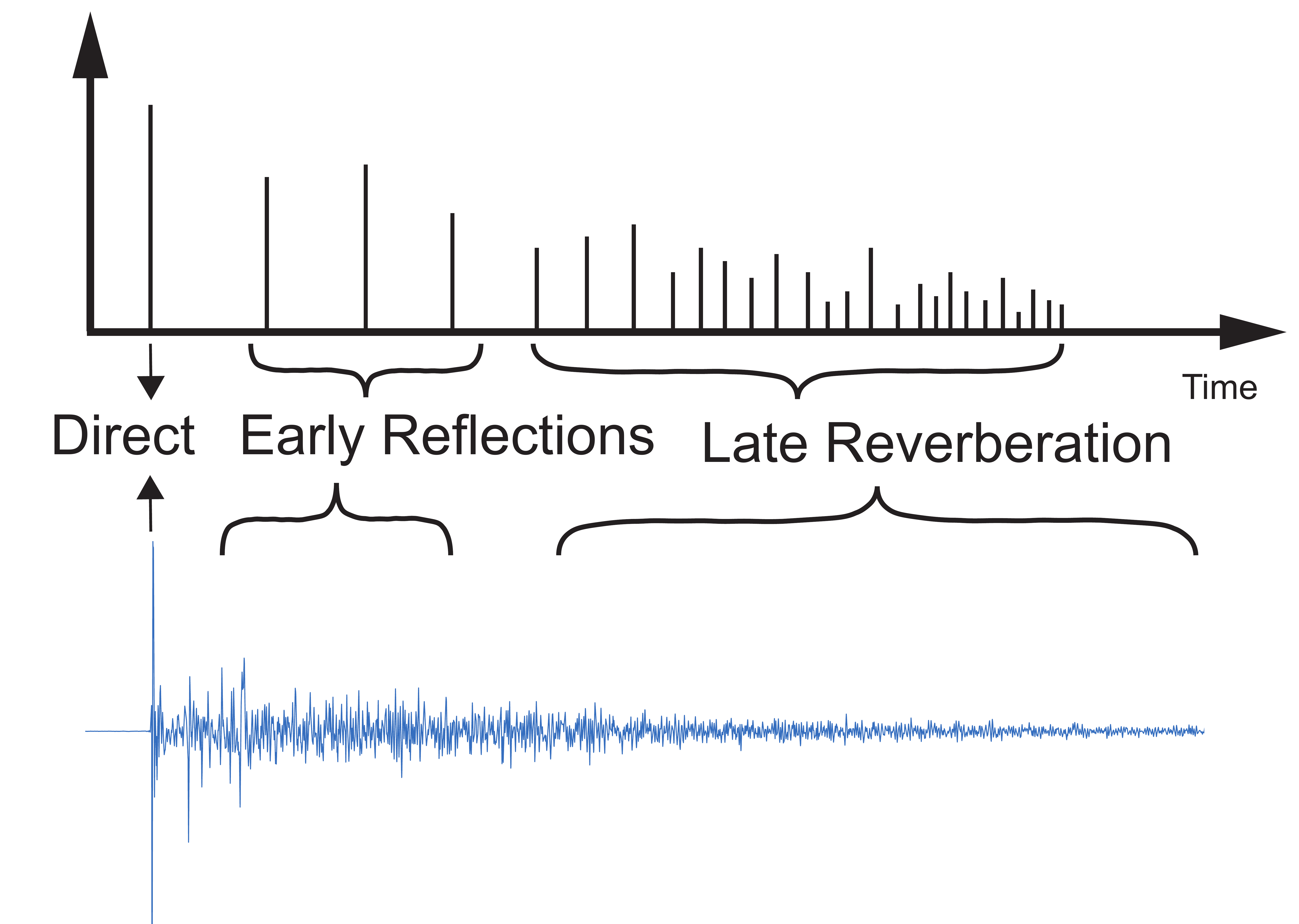}
    \vspace{-4mm}
    \caption{\textbf{A typical impulse response.} (top) An idealized illustration,
    showing the arrival time of rays and the amount of energy they carry.
    (bottom) A recorded impulse response in a lecture hall. The reflections become more dense and
    diffuse towards the later part. Traditionally, an IR is measured by recording sound omni-directionally.
    But for spatial audio generation, we need to estimate a \emph{directional}
    IR, which is illustrated in \figref{uniform_srir}.
    \label{fig:impulseParts}}
    \vspace{-2mm}
\end{figure}

\paragraph{Technical insight and contributions}
We propose to produce spatial audio by combining a lightweight measurement of room acoustics and
a fast geometric acoustic simulation.
A key step in our method is to construct \emph{directional} impulse response (IR) functions.
For traditional, non-spatial audio, an acoustic IR (see \figref{impulseParts}) is the sound
recorded omni-directionally at a listening location due to an
impulsive signal at a source location.
Then, given any input sound signals at the source, the received non-spatial sound
signals can be computed by convolving the input signals with the IR.  However,
to produce spatial audio, we need instead directional IR functions that record
the IR sound coming from each direction at the listening location.  Even in the
same scene, the IR varies with respect to the source and listening locations, and the
directional IR further depends on incoming sound directions.

An interesting property of IR functions lays the foundation for our proposed method. The
late part of the IR is the received sound energy after excessively
interacting with the scene. Every time sound waves reach a scene object, a
portion of their energy is reflected ``diffusely'', effectively causing the sound
energy distribution in the scene to become more uniform. Consequently, it is
generally accepted that the late part of the IR is independent
of the source and listening locations~\cite{Kuttruff:2017:RoomAc}.
Further, in directional IRs, the late part becomes isotropic
(independent of incoming direction), as confirmed in our room acoustic
measurements (see \figref{LRIR_direc} and \secref{validation}).

Exploiting this property, we measure a single non-spatial IR in the scene and
extract its late part through a novel method, which identifies when its energy
distribution becomes truly uniform.  This enables us to reuse the measured late
IR when constructing the spatial IR at given source and listening locations,
only relying on geometric acoustic simulation to generate the early part of the
spatial IR. The simulated early IR part is further improved by a simple and effective 
frequency modulation method that accounts for room resonances.

To leverage acoustic simulation, we reconstruct rough scene geometry from the
\tsd video footage, using a state-of-the-art \tsd structure-from-motion method,
guided by a few user specifications.  We develop an optimization approach that
estimates the acoustic material properties associated with the geometry, based
on the measured IR. The geometry and material parameters enable the acoustic
simulation to capture the early, directional component of the spatial IR.
Because the early part of the IR is oftentimes very short (typically 50-150
ms), the sound simulation is fast.

We demonstrate the quality of our resulting audio by comparing with spatial
audio directly recorded by ambisonic microphones, and show that
their differences are almost indistinguishable.
Unlike ambisonic recordings,
our method requires only a low-cost microphone, and
offers the flexibility to add, replace, and edit spatial audio for \tsd video.
We explore the potential use of our method in several applications.
While our method is designed for indoor \tsd video,
we further explore its use for those shot in outdoor spaces.

\section{Related Work}
Recent advances in \tsd video research have focused mostly on improving \emph{visual} quality.
Rich360 and Jump designed practical camera systems and developed seamless
stitching with minimal distortion, even for high-resolution \tsd
videos~\cite{lee2016rich360,anderson2016jump}.
To capture stereo omni-directional videos, Matzen et
al.~\shortcite{matzen2017low} built a novel capturing setup from off-the-shelf components,
providing a more immersive viewing experience in head-mounted displays with depth cues.
Kopf~\shortcite{kopf2016360} introduced a \tsd video stabilization algorithm for
smooth playback in the presence of camera shaking and shutter distortion.
Our work improves the audio experience in existing \tsd videos,
working in tandem with existing methods for capturing, post-processing, and playback
for immersive visual media.

Spatial audio in virtual reality (VR) is also crucial to provide convincing immersion.
Most recent work aims to enable efficient rendering of spatial audio at real-time rates.
Schissler et al.~\shortcite{schissler2016efficient} proposed a novel analytical formulation for
large area and volumetric sound sources in outdoor environments.
Constructing spatial room impulse responses (SRIR) with geometric acoustics is expensive due to
the number of rays and the disparity in energy distribution.
Schissler et al.~\shortcite{schissler2017efficient} partition the traditional IR into segments
and project each segment onto a minimal order spherical harmonics bases to
retain the perceptual quality.
We build upon the concept of SRIR and observe that late reverberation is diffuse, which means
that the late IR tail is uniform not only spatially but directionally.
Our method combines early IR simulation with estimated material parameters and
 recorded late IR tails to generate scene-aware audio for \tsd videos.

The simulation of sound propagation has been
widely studied~\cite{Vorlander:2008:Auralization,bilbao:2009:NSS}.
Wave-based methods usually provide high accuracy but require expensive
computation~\cite{Raghuvanshi:2009:ARD}.
Alternatively, geometric acoustic (GA) methods can be used, which make the
high-frequency Eikonal ray approximation~\cite{Savioja:2015:GASurvey}.  These
methods often bundle rays 
together and trace as beams for efficiency~\cite{Funkhouser:1998:BTA}. 
While traditional GA
does not include diffraction effects, they can be approximated via the uniform
theory of diffraction for edges that are
much larger than the wavelength~\cite{Tsingos:2001:UTD,Schissler:2014:HDD}.
We use the GA method proposed by Cao et al.~\shortcite{Cao:2016:BiDir}, which exploits bidirectional
path tracing and temporal coherence to provide significant speedups over
previous work.
For fast auralization in VR, many methods precompute IRs or
wavefields~\cite{Pope:1999:Realtime,Tsingos:2009:Precomputing,Raghuvanshi:2014:PWF}.
Raghuvanshi et al.~\shortcite{Raghuvanshi:2010:PWS} precompute and store one
LRIR per room, similar to our method.
We show how to use recorded IRs to optimize acoustics materials for simulation,
and also how to directly use the recorded IR tails instead of
simulating them, reducing the computation time and memory requirements.  
Moreover, our method accounts
for a particular wave effect, the room resonances, using a frequency modulation algorithm,
which further improves the generated audio quality.

To synthesize scene-aware audio, optimal material parameters are needed
in the simulation.
Given recorded IRs, we estimate the material parameters that best resemble the actual recording.
For rigid-body modal sounds, Ren et al.~\shortcite{ren2013example} optimized the material
parameters based on recordings and demonstrated the effectiveness of optimized parameters to
virtual objects.
Most related to ours is Schissler et al.~\shortcite{schissler2017acoustic}
where a pretrained neural network is used to classify the objects, followed by
an iterative optimization process.
Every iteration requires registering the simulated IR with a measured IR and solving
    a least-squares problem.
We draw inspirations from inverse image rendering problems~\cite{marschner1998inverse},
and derive an analytical gradient to the inverse material optimization problem,
which we solve in a nonlinear least-squares sense.
Our optimization runs in seconds, tens of times faster than previous work.

While our method aims to ease the audio editing process, this is a
broad area with an abundance of prior work.
Most methods strive to provide
higher-level abstractions and editing powers, to help users avoid non-intuitive
direct waveform editing.
VoCo~\cite{Jin:2017:VoCo} allows realistic text-based insertion and replacement of audio
narration using a learning-based text to speech conversion which matches
the rest of the narration.
Germain et al.~\shortcite{Germain:2016:EqMatch} present a method for equalization
matching of speech recordings, to make recordings sound as if they were recorded in
the same room, even if they weren't.
Rubin et al.~\shortcite{Rubin:2013:CTE} present an interface for editing audio
stories like interviews and speeches, which includes
transcript-based speech editing, music browsing, and music retargeting.
Like previous work, we aim to match the timbre of generated sounds with that of recordings.
Moreover, we are able to produce spatial audio that blends in seamlessly with
existing \tsd videos, and provide a high-level ``geometric'' effect which can be
applied to audio.

\section{Rationale and Overview}
An important concept used throughout our method is the acoustic impulse
response (IR). We therefore start by discussing its properties in
typical indoor scenes to motivate our algorithmic choices. 

\subsection{Properties of Room Acoustic Impulse Response}\label{sec:properties}
The room acoustic IR is a time-dependent function, describing the sound signals
recorded at a listening location due to an impulsive (Dirac delta-like) signal at a source
(\figref{impulseParts}).  In this paper, we use $H(t)$ to denote an IR.
If $H(t)$ is known, then the sound signal $s_\txt{r}(t)$ received
at the listening location can be computed by convolving $H(t)$ with the sound
signal $s_\txt{e}(t)$ emitted from the source:
$s_\txt{r}(t) = s_\txt{e}(t)*H(t)$.
Therefore, to add spatial audio to a \tsd video, we need to estimate the IRs
between the sound source and the camera location in the scene along all
incoming directions.

The IR is usually split into three parts: \textit{i)} the direct sound traveling from the
source to the listener, \textit{ii)} the first few early reflections (ER), and \textit{iii)}
the later reflections called late reverberation (LR).
Part (i) and (ii) are the early reflection impulse response (ERIR). Perceptually, they
give us a directional sense of the sound source, known as the precedence effect~\cite{Gardner:1968:Precedence}.

The LR part of the impulse response (referred to as LRIR) has several properties
significant to our goal. First, the LRIR is greatly ``diffused'' in the
scene~\cite{Kuttruff:2017:RoomAc}, meaning that it has little dependence with respect to
the source and listening locations. This is because whenever a directional sound wave
encounters an obstacle, a portion of the energy is reflected diffusely,
spreading the sound in many directions.
Virtually all rooms include some diffuse reflection
even when the walls appear smooth~\cite{Hodgson:1991:Diffuse}.
Thus, the longer the sound travels in a scene, the more it gets diffused.
The LRIR has little perceptual contribution to our sense of
directionality. Rather, it conveys a sense of ``spaciousness''~\cite{Kendall:1995:Decorr} --- the size
of the room, but not where the listener and source are.

\begin{figure}[t]
    \includegraphics[width=0.99\columnwidth]{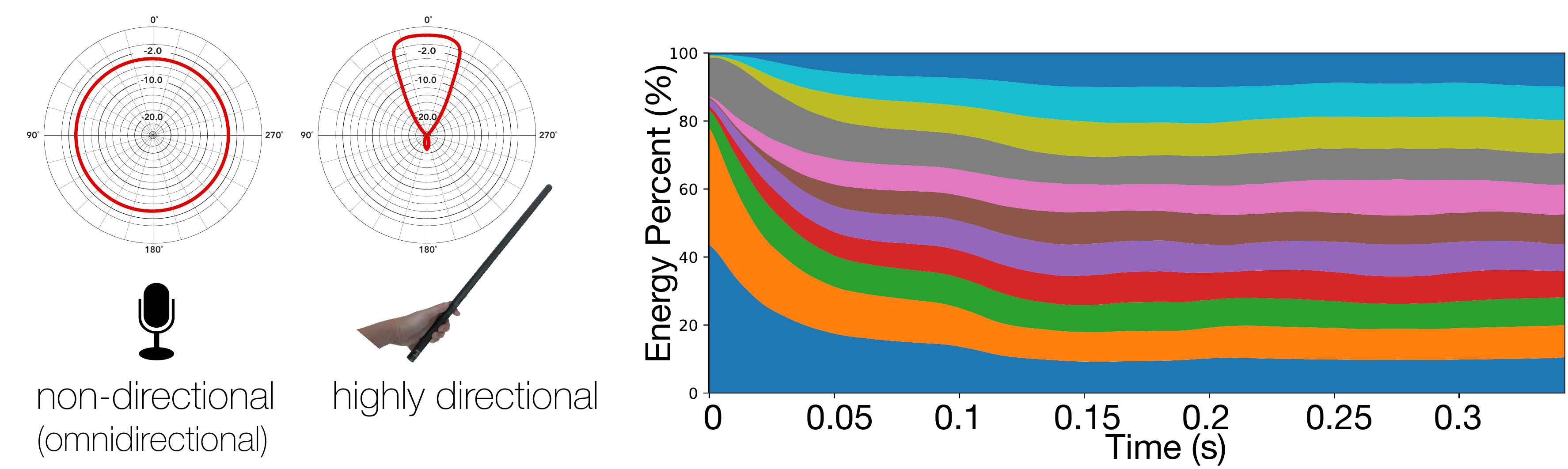}
    \vspace{-3mm}
    \caption{{\bf LRIR isotropy.}
    We use a high-end directional (shotgun) microphone (RODE NTG8) to measure room acoustic IR received
    along particular directions.
    (left) The polar pickup plot of our shotgun microphone in comparison to
    the conventional omni-directional microphone. The shotgun microphone records
    sound mainly from its front direction.
    (right) For several recordings of an impulse with the
    shotgun microphone pointed in different directions (corresponding to different
    colors), we plot the amount of energy coming from each direction with respect to time.
    In the early part, more energy is in directions that face the source, but the
    energy is quickly distributed uniformly among all directions.
    \label{fig:LRIR_direc}}
    \vspace{-6mm}
\end{figure}

Another important implication of LRIR being diffused is that the sound energy carried by LRIR
tends to be uniformly distributed, not only spatially~\cite{Kuttruff:2017:RoomAc}
but also directionally --- it can be assumed isotropic.
We justify this assumption with directional acoustic measurements, as
described in \figref{LRIR_direc}.

\subsection{Method Overview}\label{sec:overview}
The room acoustic IR properties suggests a hybrid approach for estimating
spatial IRs when we generate spatial audio for \tsd videos: the LRIR can be
measured at one pair of source and listening locations because of its spatial
and directional independence, while the ERIR needs to be simulated with
carefully chosen parameters to capture sound directionality.
Also crucial is the time duration
for separating ER from LR, in order to ensure the directional independence of LRIR
satisfied. 
The major steps of our method are summarized as follows.

\begingroup
\setlength{\columnsep}{0pt}%
\textit{\tsd video analysis.}\, 
Provided a \tsd video, we estimate rough scene geometry and the camera
trajectory in the scene. The former is for running the simulation, and the latter
is to locate the listener when we generate spatial audio.
3D scene reconstruction has been an active research area in computer vision. 
We adopt the recent
\begin{wrapfigure}[11]{r}{0.33\columnwidth}
  \vspace{-5mm}
    \includegraphics[width=0.14\textwidth]{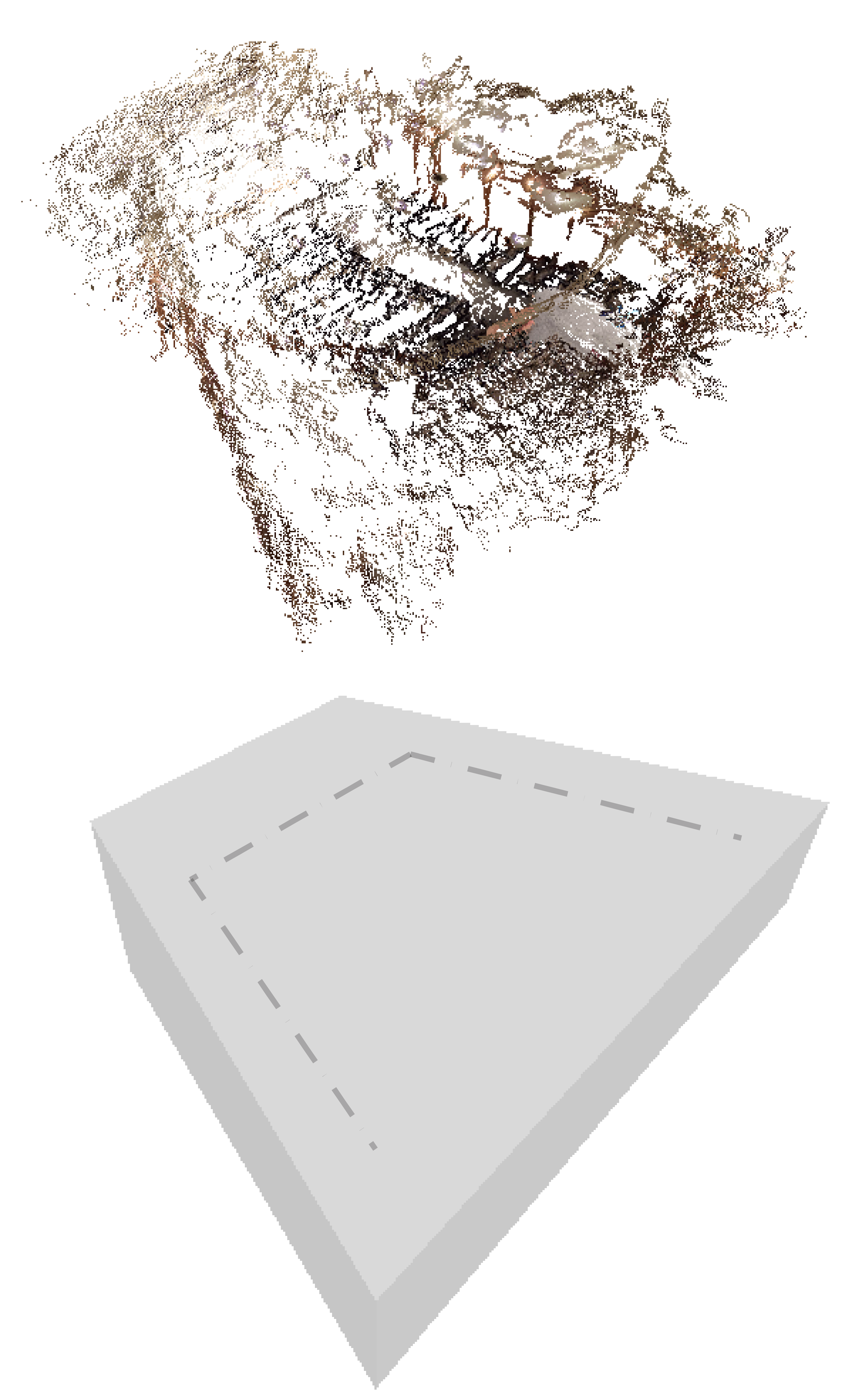}
\end{wrapfigure}
 structure-from-motion approach~\cite{Huang:2017:6DOF} that
generates a point cloud of the scene from a \tsd video (see top of the adjacent figure), 
along with an aligned camera path through it.
Our method does not depend on this particular approach: any future improvement
could be easily incorporated.
Then, we rely on the user to specify a few planar shapes that align with the point cloud
to capture the main geometry of the scene, such as the roof, floor, and walls (see bottom
of the adjacent figure).
A benefit of our hybrid
approach is that only approximate geometry is required: it does not need to be water-tight, or even manifold.
The method of~\cite{Huang:2017:6DOF} takes 10-20 minutes, depending on the video resolution.
Creating planar geometry to match the reconstructed point cloud only takes
users several minutes per room.

\textit{Room IR analysis.}\,
Next, we record an impulse response in the room using
a conventional omni-directional microphone and speaker (\secref{irMeas}).
This measurement is straightforward and serves two purposes. First, it provides the LR
component when we estimate the spatial IR between a sound source and a camera
location.  Second, it offers a means to sense the acoustic material properties
in the room. Based on the measured IR, we estimate the acoustic material
parameters for use in acoustic simulation, by formulating
a nonlinear least-squares optimization problem (\secref{mat_ana}).
After acquiring scene material parameters, we are then able to leverage the acoustic
simulation to determine the transition point between the ERIR and LRIR
based on a \emph{directionality} analysis of the incoming sound energies (\secref{erDuration}).

\textit{Spatial audio generation.}\,
Lastly, we generate spatial \tsd audio from input audio signals.
As audio editors place sources in the scene, our simulator computes the ERIR
from the source to positions on the reconstructed camera path (providing
directional cues), and the LRIR is reused
from the measured IR (providing a sense of spaciousness).
Combining them together, we obtain spatial IRs for generating
spatial audio (\secref{runtime}).
\camera{We will show how to store the final spatial audio in ambisonics 
    (in \secref{ambisonics}), which can be encapsulated in the standard \tsd video format
to adapt sound effects to the view direction when the video is played back.}
\endgroup

\subsection{Room Acoustic Simulation}\label{sec:sim}
Before diving into our technical details, we briefly describe the acoustic
simulator that we use. We use a geometric acoustic (GA) model that describes
sound propagation using \emph{paths} along which sound energy propagates
from the source to the receiver,
akin to the propagation of light rays through an environment.
Each path carries a certain amount of sound energy, and arrives at the receiver
with a time delay proportional to the path length.
Exploiting the sound energy carried by the paths and their arrival time, we are
able to infer scene materials (\secref{mat_ana}), determine ER duration
(\secref{erDuration}), and synthesize ERIRs for ambisonic audio generation
(\secref{direction_aware}).

Our method does not depend on any particular GA method. In this paper, we
employ the bidirectional path tracing method recently developed
in~\cite{Cao:2016:BiDir}. This technique simulates sound propagation by
tracing paths from both the sound source and the receiver, and uses multiple
importance sampling to connect the forward and backward paths.
It offers a considerable speedup over prior GA algorithms and better balance
between early and late acoustic responses.

While the GA model is an approximation of sound propagation and ignoring wave
behaviors such as diffraction, it can reasonably estimate the impulse response
of room acoustics, and has been widely used for decades~\cite{Savioja:2015:GASurvey}.
Nevertheless, we consider an important wave effect, namely \emph{room resonance}, and propose
a frequency modulation method to incorporate the room resonance effect in our simulated ERIR (\secref{modulation}).

\section{Room Acoustic Analysis for \tsd Scenes}\label{sec:audioAnalysis}
This section presents our method of analyzing an IR measurement to estimate
acoustic material properties of the room and frequency modulation coefficients
needed for compensating room resonances.
We also determine the transition point between ERIR and LRIR.

\subsection{IR Measurement}\label{sec:irMeas}
\begin{figure}[t]
 \centering
 \includegraphics[width=0.99\columnwidth]{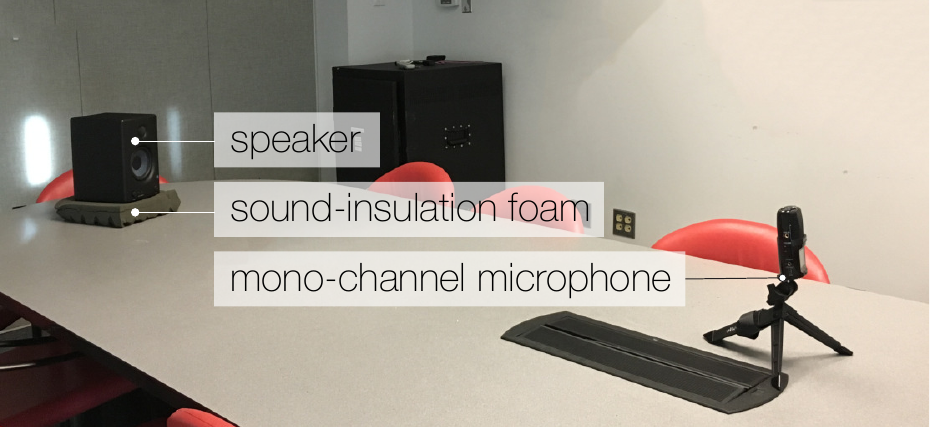}
 \vspace{-3mm}
 \caption{{\bf IR measurement.} We measure the IR using a conventional speaker and a mono-channel microphone.
 The speaker plays a sine sweep noise, which is then recorded by a microphone.
 In practice, we put the speaker on soft foam to absorb any mechanical
 vibrations it produces, which
 can be propagated to the microphone through the table.
 \label{fig:measurement}}
 \vspace{-2mm}
\end{figure}

There exist many methods for acoustic IR measurement.
In this work, we use the reliable sine sweep
technique of~\cite{farina2000simultaneous,farina2007advancements}.
We briefly summarize its theoretical foundation here:
the signal $s_\text{r}(t)$ recorded by a receiver is the convolution of the source
signal $s_\text{e}(t)$
and the room's IR $H(t)$ (i.e., $s_\text{r}(t) = s_\text{e}(t)*H(t)$).
It can be shown that $H(t)$ can be reconstructed by measuring the
cross-correlation between $s_\text{r}(t)$ and $s_\text{e}(t)$, $H(t) =
s_\text{r}(t)\star s_\text{e}(t)$, as long
as the autocorrelation of the source signals $s_\text{e}(t)$ is a Dirac delta, or
$s_\text{e}(t)\star s_\text{e}(t)=\delta(t)$.
For reliability, $s_\text{e}(t)$ needs to have a flat power spectrum.
A commonly used practical choice of $s_\text{e}(t)$ is a sine sweep function that
exponentially increases in frequency from $\omega_1$ to $\omega_2$ in a
time period T~\cite{farina2000simultaneous}:
\begin{equation}
    s_\text{e}(t) = \sin\left[ \frac{\omega_1
    T}{\ln\frac{\omega_2}{\omega_1}}\left(e^{\frac{t}{T}\ln\frac{\omega_2}{\omega_1}}-1\right)\right].
\end{equation}
This signal spends more time sweeping the low-frequency regime, thus it is
particularly robust to low-pass noise sources like those in most rooms.
In practice, we choose $\omega_1=20$ Hz, $\omega_2= 20$ kHz, and $T=48$ seconds.
Also, we play the source $s_\text{e}(t)$ and record $s_\text{r}(t)$ simultaneously, so they are fully
synchronized. This sine sweep is played only once (no average is needed),
using a conventional speaker and a mono-channel microphone.
Their simple setup is illustrated in Figure~\ref{fig:measurement}.

While the IR depends on the positions of source and receiver, our measurement
is insensitive to where the source and receiver are positioned. This is because
for ambisonic audio generation, we only need the LR component of the measured IR,
which remains largely constant in the environment (recall \secref{properties}).
In practice, we position
the source and receiver almost arbitrarily, as long as they are well separated.
We only need to perform the IR measurement once in a room.
If there are multiple rooms in the \tsd scene, we measure one IR per room (an
example is shown in \secref{app}).
This step yields a measured IR, $H(t)$, and we also compute its energy response
$h(t)$.  

\subsection{Material Analysis}\label{sec:mat_ana}
Having the IR measured and the room's rough geometry reconstructed from
the \tsd video,
we now determine the acoustic material parameters needed for our room acoustic simulation.
These parameters are associated with individual planar regions of
the reconstructed room shape --- for example, in a typical room, the walls are
often painted with a particular acoustic material while the floor may have other
acoustic properties.
Our method also allows the user to manually select sections of the reconstructed
geometry and group them as having the same acoustic material.

Acoustic properties of materials are frequency dependent. We therefore define these acoustic
parameters in each octave frequency band.
Without loss of generality, consider a particular octave band.
When a sound wave in this octave band is reflected by a material $i$,
part of the sound energy is absorbed by the material, which is described
by the material absorption coefficient $p_i$ in this octave band.
Let $\vec{p}$ stack the $p_i$ values of all types of materials in the room.
We then formulate an optimization problem to solve for $\vec{p}$.

\begin{figure}[t]
 \centering
 \includegraphics[width=0.8\columnwidth]{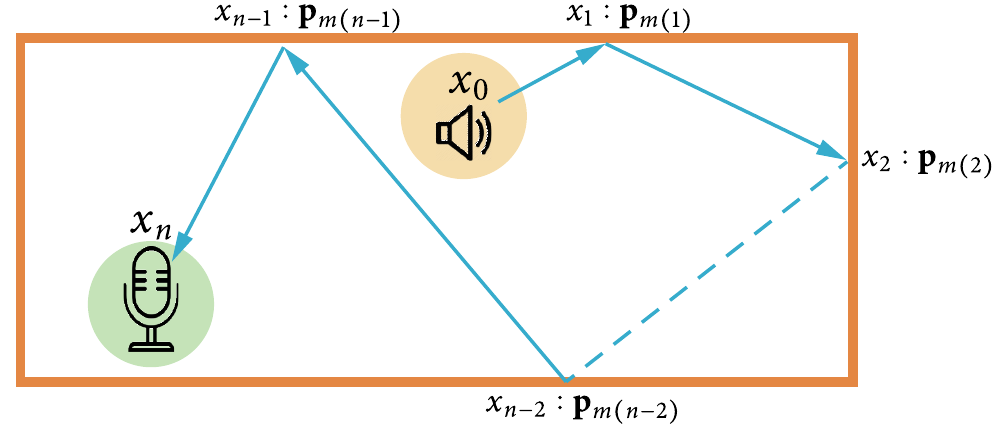}
 \vspace{-2mm}
 \caption{{\bf Path and notation.} 
 A sound path connecting a source to a receiver may be reflected multiple times
 at the surface positions $\bm{x}_i$. Each $\bm{x}_i$ is associated with a material
 indexed by $m(i)$, and its absorption coefficients over all frequency bands are stacked
 in the vector $\vec{p}_{m(i)}$.
 \label{fig:path}}
 \vspace{-6mm}
\end{figure}

\textit{Path.}\,
The ray-based room acoustic simulator generates numerous paths, along which sound signals
propagate from a source to a speaker.
Each path is described by a sequence of 3D positions, $\vec{x}_0,\vec{x}_1,\mdots,\vec{x}_n$,
where the first and last positions are the source and receiver, respectively. The other positions
are surface points where the ray is reflected, each associated with an acoustic
material (\figref{path}).
Depending on the material at position $\vec{x}_i$, $i=1\mdots n-1$, each
$\vec{x}_i$ is mapped to an absorption coefficient indexed in the aforementioned
parameter vector $\vec{p}$. Let $m(i)$ denote the index.

\textit{Energy.}\,
With this notion, the energy fraction propagated along a path $j$ and arriving at
the receiver is written as
\begin{equation}\label{eq:Ej}
    e_j(\vec{p}) = \beta_j \prod_{i=1}^{N_j} \vec{p}_{m(i)},
\end{equation}
where $N_j$ is the number of surface reflection points along the path $j$,
and $\beta_j$ accounts for the sound attenuation due to propagation in air; it depends
on the path length but not on room materials~\cite{dunn2015springer}.
Our goal is to determine $\vec{p}$ so that the energies $e_j$ delivered by all paths
at the receiver match the energy distribution in the measured IR.

\textit{Objective function.}\,
To this end, we propose the following nonlinear least-squares objective function,
\begin{equation}\label{eq:opt}
    J(\vec{p}) = \sum_{j=1}^M \left[ \log_{10}\left( \frac{e_j(\vec{p})}{e_0}\right) - \log_{10}\left( \frac{\tilde{h}(t_j)}{\tilde{h}(\bar{t}_0)} \right)\right]^2,
\end{equation}
where $j\in[0,M]$ is the index of the paths resulted from the simulation,
$t_j$ is the sound travel time along path $j$,
$\bar{t}_0$ is the earliest sound arrival time in the measured IR (not in the simulation),
and $\tilde{h}(t_j)$ is a parametric model of the measured sound energy response at time $t_j$,
which we will elaborate on shortly.

Moreover, $e_0$ is the energy delivered by the earliest path
arriving at the receiver in the simulation.
This is the path that directly connects the source and receiver, thus independent from
material parameters.
This is also the path whose arrival time
is used to calibrate the reconstructed room size: before formulating the objective function~\eq{opt},
we scale the room size so that the arrival time of the first path matches
$\bar{t}_0$, and in turn, the same scale is applied to the arrival time $t_j$
of all later paths.
By taking the ratio of $e_j$ to $e_0$, we avoid
matching the absolute energy level between the simulation and the measurement.

We use a fitted parametric model $\tilde{h}(t)$ in~\eq{opt} instead of the measured
energy response $h(t)$, because using $h(t)$ is susceptible to measurement noise.
Traer and McDermott~\shortcite{Traer:2016:ReverbStat} measured the IRs of hundreds of
different daily scenes, and discovered that $h(t)$  decays exponentially, and the
decay rates are consistently frequency dependent. Thus, we fit the measured $h(t)$ in
each frequency band $j$ with an exponentially decaying function, $\tilde{h}_j(t) = A_j e^{-\gamma_j t}$,
and use it in~\eq{opt}, where we discard the subscript $j$ for simplicity, as \eqnref{opt}
is solved for each frequency band independently.

We note that it is critical to formulate the objective function~\eq{opt}
using a logarithmic scale, because the ray energy drops exponentially with respect to
the arrival time. Otherwise, the summation in the nonlinear least-squares sense would overemphasize
the match of the early paths while sacrificing late paths, which also have significant
perceptual contributions~\cite{Traer:2016:ReverbStat}.

\paragraph{Inverse Solve}
\camera{We solve for $\vec{p}$ by minimizing~\eq{opt} with the constraint that all values in $\vec{p}$
must  lie in $[0, 1]$} . This constrained nonlinear least-squares problem can be efficiently solved
using the L-BFGS-B algorithm~\cite{Zhu:1997:ALF}. This is a
gradient-descent-based method, where the
gradient of~\eq{opt} is
\begin{equation}\label{eq:der}
    \frac{\partial J}{\partial p_i} = \frac{2}{\ln10}\sum_{j=1}^M \frac{1}{e_j}\left[ \log_{10}\left( \frac{e_j}{e_0}\right) - \log_{10}\left( \frac{\tilde{h}(t_j)}{\tilde{h}(\bar{t}_0)} \right)\right]\frac{\partial e_j}{\partial p_i}.
\end{equation}
In practice, the optimizations for individual frequency bands are performed in
parallel, and often take less than $10$ seconds.  

\begin{figure}[t]
 \centering
 \includegraphics[width=0.99\columnwidth]{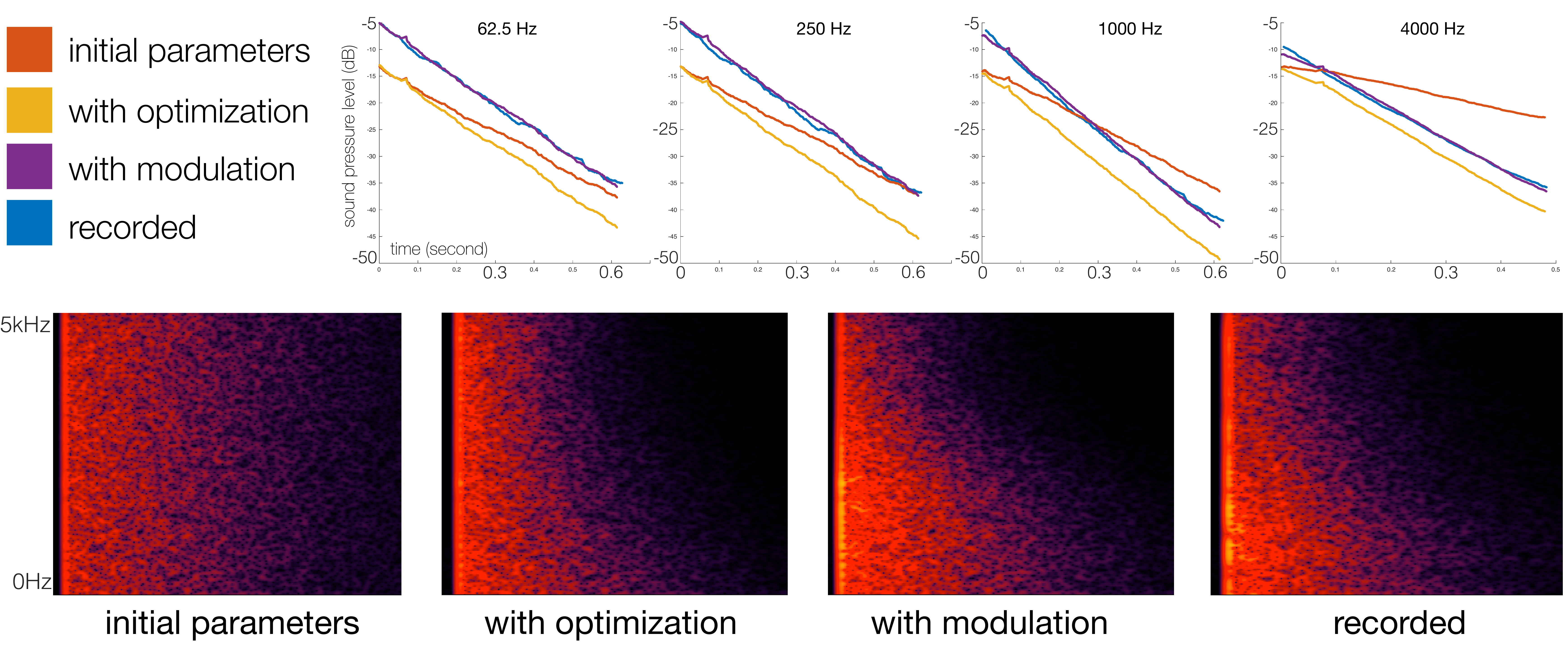}
 \vspace{-2mm}
 \caption{{\bf IR optimization.} 
     (top) The four plots correspond to four frequency bands (centered at 62.5Hz, 250Hz, 1000Hz, and 4000Hz).
     In each plot, the four curves correspond to the energy decay curves of four IRs obtained using different approaches.
     The orange curves are simulated using initial material parameters, and
     serve as a starting point.
     The blue target curves are directly recorded.
     The yellow curves are simulated using our optimized material parameters. They have the same energy decay rates
     as the measured (blue) curves but different scales.
     The purple curves are computed using the yellow curves modified using our frequency modulation algorithm 
     (see \secref{modulation} and Eq.~\ref{eq:ht}), and they match the measured curves closely.
     The spectrograms of the four IRs are shown on the bottom, where the simulated IR with frequency modulation matches
     closely with the recorded IR.
 \label{fig:opt-results}}
 \vspace{-5mm}
\end{figure}

As a validation, we substitute the optimized material absorption coefficient
$\vec{p}$ into~\eq{Ej}, and evaluate the energy $e_j$ of every path $j$ we collected.
Using these updated $e_j$, we construct a simulated IR and compare it with the measured IR.
As shown in the top row of \figref{opt-results}, the energy decay rate of the \emph{simulated} IR with
respect to time indeed matches with the measured IR at every frequency band.
This verifies the plausibility of our optimized parameter values.
Nevertheless, the energy intensities are still different.
It is this discrepancy that motivates our frequency modulation analysis, as described next.

\subsection{Frequency Modulation Analysis}\label{sec:modulation}
In our simulated IR, the energy decay in every frequency band always starts
from $e_0$, the energy level delivered by the direct path from the source to
the receiver. This is because the direct path has no surface reflection,
and is thus independent of the material's absorption. However, this reasoning contradicts
what we observe in the measured IR, where the energy decay in different
frequency bands start from different values (e.g., see the four dark blue
curves in the top row plots of \figref{opt-results}).
An important factor that cause the frequency-dependent variation is
a wave behavior of sound.
namely the \emph{room resonances}. In essence, each room is an acoustic chamber.
When a sound wave travels in the chamber, it boosts wave components
at its resonant frequencies while suppressing others.
Most rooms have their fundamental resonances in the 20-200Hz range. It is known
that the room resonances affect the sound effects in the room and are one of
the major obstacles for accurate sound reproduction~\cite{cox2004room}.  Yet,
room resonance, because of its wave nature, cannot be captured by a
geometric acoustic (GA) simulation.

We propose a simple and effective method to incorporate room resonances in our simulated IR.
We use $\tilde{H}(t)$ to denote our simulated IR and to distinguish from the measured IR $H(t)$.
Let $t_0$ be the arrival time of the direct path. We compute the discrete Fourier transforms
of the simulated and measured IRs in a small time window $\Delta t$ at $t_0$:
\begin{equation}
    \tilde{\mathsf{H}}(\omega) = \mathcal{F}[\tilde{H}(t)]\;\textrm{and}\;
    \mathsf{H}(\omega) = \mathcal{F}[H(t)],\;\textrm{for } t_0<t<t_0+\Delta t.
\end{equation}
Both $\tilde{\mathsf{H}}(\omega)$ and $\mathsf{H}(\omega)$ in the discrete setting are vectors of
complex numbers. We compute and store the ratio $\mathsf{M}(\omega) =
{|\mathsf{H}(\omega)|}/{\tilde{|\mathsf{H}}(\omega)|}$. Later, when we generate spatial IRs,
we will use $\mathsf{M}(\omega)$ to modulate the frequency-domain energy of the IRs without affecting
their phases (see \secref{direction_aware}).

\begin{figure}[b]
 \vspace{-3mm}
\includegraphics[width=0.8\hsize]{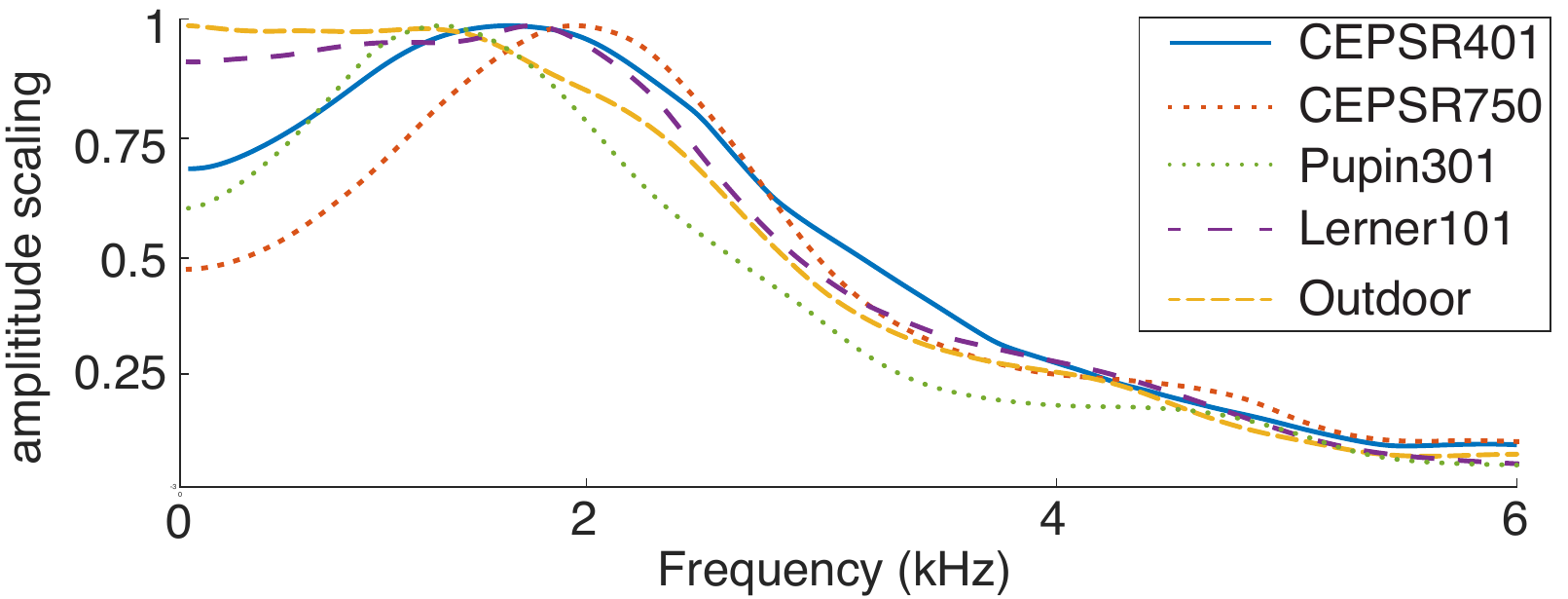}
\vspace{-3mm}
\caption{\camera{\textbf{Frequency modulation profiles.} 
We plot the amplitude ratio $\mathsf{M}(\omega)$ measured in different rooms.
It is interesting to observe the peaks of each curve, which reflect the room
resonant frequencies. The larger the room is, the lower resonant frequency it has.
\label{fig:roomFreqProfiles}}}
 \vspace{-2mm}
 \end{figure}

In practice, we need to smooth $\mathsf{M}(\omega)$ in presence of measurement noise.
According to the uncertainty principle of signal processing~\cite{papoulis1977signal},
we choose a small window that contains 256 samples of $\tilde{H}(t)$ and $H(t)$. This gives us 128 samples
of $\mathsf{M}(\omega)$ in frequency domain. We slide the small time window
$\Delta t$ in a slightly larger window $[t_0,t_0+2\Delta t]$, repeat the
computation of $\mathsf{M}(t)$, and then average the resulting ratios.
\camera{
\figref{roomFreqProfiles} shows typical profiles of different rooms in our experiments.
}

To our knowledge, methods that compensate room resonances remain elusive in
existing GA-based audio generation approaches. As shown in \figref{opt-results}
and our supplemental video, our frequency modulation method improves the fidelity of the
simulated IR and the realism of resulting spatial audio in a very noticeable way.

\subsection{ER Duration Analysis}\label{sec:erDuration}
After obtaining the optimized material parameters,
we now use simulation to obtain a reliable estimate of the ER duration $T_\txt{ER}$.

ER-LR separation is traditionally defined based on \emph{subjective}
perception~\cite{Kuttruff:2017:RoomAc}.
There exist various heuristics for estimating the ER duration $T_\txt{ER}$ from
IR measurement or simulation, from a simple kurtosis threshold~\cite{Traer:2016:ReverbStat}
to a threshold on the number of peaks per second in a
simulated IR~\cite{Raghuvanshi:2010:PWS},
None of these heuristics rest on the
observation that we exploit to combine a simulated ERIR with a measured LRIR for
{ambisonic} audio generation --- that is, the LR is isotropic, having uniformly
distributed incoming sound energy along all directions.
As a consequence, simple heuristics lead to unreliable $T_\txt{ER}$ estimates. We therefore
propose a new algorithm to determine $T_\txt{ER}$ directly based on the
observation of the LR's isotropy.

Reusing the path energies $e_j$ collected in~\secref{mat_ana},
we define the ER duration $T_\txt{ER}$ as the earliest time instant when the received acoustic energy
is \emph{uniformly distributed} among all directions.
To identify $T_\txt{ER}$, the collected rays with their energies are viewed as Monte-Carlo
samples of the energy distribution over time and direction.
From this vantage point,
we consider a sliding time window $\Delta t$, and check if the
\emph{statistical distance} between
the energy distribution sampled by the rays in the time window and a uniform distribution is
below a threshold.

Three statistical distance metrics are commonly used, including Kolmogorov-Smirnov (KS) Distance,
the Earth Mover’s Distance, and the Cram\'{e}r-von Mises Distance.
They can be viewed as taking different kinds of norms of the
cumulative distribution function (CDF) difference between two distributions.
Here we choose to use KS distance, while the other two can be naturally used as well.

Our algorithm is as follows. 
Consider all the rays in a time window $\Delta t$=10ms.
The ray directions are described by two coordinates, the azimuthal and zenith angles.
We process the distribution with respect to each coordinate separately.
First, we put the sampled energies $e_j$ into histogram bins according to their zenith angles.
After normalization, this histogram represents a discrete probability distribution
of incoming sound energies with respect to zenith angle. We then convert this histogram into
a discrete CDF, represented by a vector $\vec{P}_s$.
If the energy is uniformly distributed, the expected CDF with respect to the zenith angle $\phi$ is
\begin{equation}
    P_c(\phi) = \frac{1}{2}(1-\cos\phi),
\end{equation}
which is discretized into a vector $\vec{P}_c$ with the same length as $\vec{P}_s$.
The KS distance is computed as $d_{\phi} = |\vec{P}_c - \vec{P}_s|_\infty$.
Similarly, we compute the KS distance $d_{\theta}$ of the energy distribution with respect to
the azimuthal angle $\theta$. In this dimension, the expected CDF is simply a linear function,
as $\theta$ needs to be uniformly distributed in $[0,2\pi]$.
If both KS distances are smaller than a threshold (0.15 in all our examples),
we consider the current sliding time window
$\Delta t$ as having uniformly distributed directional energies.
As we slide the time window, the first distance that passes the KS test determines $T_\txt{ER}$.

To verify the robustness of this method, we run the acoustic simulation seven times, each 
set to produce a different number of total rays --- the total number of rays increases
from 15000 to 38000 evenly. After each simulation, we repeat the aforementioned 
analysis to compute $T_\txt{ER}$. We verify that among all the $T_\txt{ER}$ values,
the variance is small: less than 4.2\% of the average $T_{ER}$.

\begin{figure}[t]
 \centering
 \includegraphics[width=0.75\columnwidth]{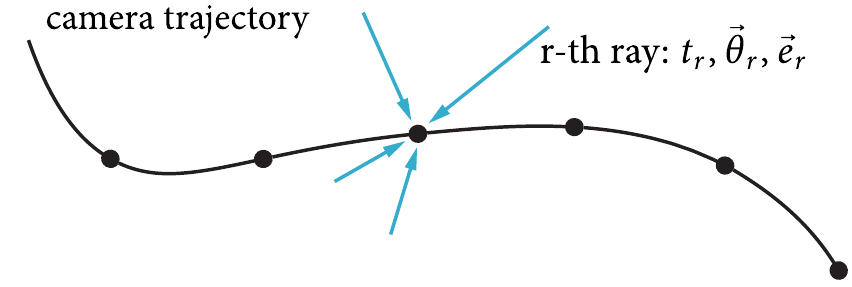}
 \vspace{-3mm}
 \caption{{\bf Ray samples.} We sample locations along the camera trajectory, and use geometric
 acoustic simulation to collect sound rays that arrive to each location within $T_\txt{ER}$ after
 an impulsive sound signal is emitted from the source.
 These rays will be used for synthesizing the ERIR for spatial audio. 
 \label{fig:ER_rays}}
 \vspace{-3mm}
\end{figure}
\section{Ambisonic Audio for \tsd Videos}\label{sec:runtime}
After analyzing the room geometry and acoustics, we are now able to generate ambisonic audio for
any \tsd video captured in the same scene.
This section describes our method which produces ambisonic audio from a
dry audio signal.  This technique will be the cornerstone of various applications that we will explore
in~\secref{app}.

\subsection{Constructing Direction-Aware Impulse Responses}\label{sec:direction_aware}
\paragraph{Trajectory analysis}
Provided a \tsd video, we recover the camera motion path by
performing structure-from-motion analysis~\cite{Huang:2017:6DOF}.
This is the same technique that we use for reconstructing the room shape
(recall~\secref{overview}). Our method does not critically depend on this
technique; any source of geometry and a registered camera trajectory would
suffice.

\paragraph{Simulating ER}
To add ambisonic sound to a \tsd video, the user first clicks a location in the
reconstructed 3D scene to specify a sound source position.
The source location, the camera trajectory, and the room geometry together with the optimized acoustic
materials provide sufficient information to launch a room acoustic simulation.
The goal of this simulation is to collect a set of incoming acoustic rays at each sampled location along
the camera trajectory. These rays will be used to construct directional IRs for early reverberation.
Therefore, in our path-tracing acoustic simulation, we cull a path whenever its travel time exceeds $T_\txt{ER}$.
This restriction of simulating only early paths significantly lowers the
simulation cost.  In our implementation, culling paths using $T_\txt{ER}$ yields 10$\sim$20$\times$ speedups and memory savings
in comparison to a simulation that lasts for the time length of measured IR.

In practice, we sample positions every 50 centimeters along the camera trajectory,
and for each position, we collect incoming rays that arrived before
$T_\txt{ER}$. Each ray is described by its arrival time, its incoming direction $\vec{\theta}$
(including azimuthal and zenith angles) and the carried sound energy $e_i$ of every octave band $i$ (see \figref{ER_rays}).

\paragraph{Constructing IRs}
Next, at every camera position, we construct spatial IRs for ambisonic audio synthesis.
Each spatial IR is decomposed into two components. The early reverberation component (ERIR) is directional,
constructed individually from the simulated early rays.
Given a ray $r$ coming from the direction $\vec{\theta}$ and carrying energies
$e_{r,i}$ of all octave bands (index by $i$), we generate an ERIR component
$H^*_{r,\vec{\theta}}(t)$ using the classic Linkwitz-Riley 4th-order crossover
filter, as was used in~\cite{Schissler:2014:HDD}.

At this point, we apply the frequency modulation curve $\mathsf{M}(\omega)$ that we computed
in \secref{modulation} to $H^*_{r,\vec{\theta}}(t)$, because the early rays
resulting from
GA-based simulation do not capture the room resonances.
In particular, we compute the Fourier transform of $H^*_{r,\vec{\theta}}(t)$ to get
$\mathsf{H}^*_{r,\vec{\theta}}(\omega) = \mathcal{F}[H^*_{r,\vec{\theta}}(t)]$, and scale it
using $\mathsf{M}(\omega)$ before transforming it back in time domain. The resulting ERIR,
\begin{equation}\label{eq:ht}
H_{r,\vec{\theta}}(t) = \mathcal{F}^{-1}[\mathsf{H}^*_{r,\vec{\theta}}(\omega)\mathsf{M}(\omega)],
\end{equation}
is what we will use for spatial audio generation (\secref{ambisonics}). As shown in the supplemental video's
soundtrack, this step improves the realism of resulting spatial audio in a noticeable way.

The LRIR component $H_\txt{L}(t)$ is omni-directional, directly taken by scaling the measured IR $H(t)$ for $t>t_\txt{ER}$:
\begin{equation*}
    H_\txt{L}(t) = \begin{cases}
        0,  &  t < t_\txt{ER}\\
        \left[\left(\int_{T_\txt{ER}-\Delta t}^{T_\txt{ER}} h(t) \txt{d}t\right)^{-1}\sum_{r\in\mathcal{W}}\sum_i e_{r,i}\right]^{\frac{1}{2}} H(t), & t\ge t_\txt{ER},
    \end{cases}
\end{equation*}
The scale in front of $H(t)$ is to match the energy level when combining
simulated ERIR with the measured LRIR.
It ensures that,
in a small time window $\Delta t$ near $T_\txt{ER}$,
the ratio of ERIR energy to LRIR energy in the synthesized IR is
the same as the ratio computed using the measured energy response $h(t)$. Here,
$\mathcal{W}$ denotes the set of rays whose arrival time
is in the time window $[T_\txt{ER}-\Delta t, T_\txt{ER}]$, and the index $i$ in
the summation iterates through all octave bands.

\subsection{Generating Ambisonic Audio}\label{sec:ambisonics}
Lastly, provided a dry audio, we generate ambisonic audio received as the camera moves along its trajectory.

\textit{Background.}
Ambisonic audio uses multiple channels to reproduce the sound field arriving to a receiver from all directions.
It can be understood as an approximation to the solution of the nonhomogeneous Helmholtz equation,
\begin{equation}\label{eq:hel}
    (\Delta + k^2) p = -f_k(\vec{\psi})\frac{\delta(r-r_\txt{L})}{r^2_\txt{L}},
\end{equation}
for each frequency band~\cite{zotter2009alternative},
where $p$ is the received sound pressure, $k$ is the wave number of the frequency band,
$r_\txt{L}$ is the distance of sound sources from the receiver,
and $f_k(\vec{\psi})$ is the directional distribution of the sound sources at the frequency band $k$.
In our case, at each location along the camera trajectory,
$f_k(\vec{\psi})$ is specified by its incoming rays.
If a receiver is located at a polar coordinate $(r,\vec{\psi})$,
then its sound pressure is described by the solution of~\eq{hel},
\begin{equation}\label{eq:decode}
    p_k(r,\vec{\psi}) = -ik\sum_{n=0}^\infty\sum_{m=-n}^n\phi_{k,nm}Y_n^m(\vec{\psi})h_n(kr_\txt{L})j_n(kr),
\end{equation}
where $Y_n^m(\vec{\psi})$ are the real-valued spherical harmonics,
$j_n$ are the spherical Bessel functions, $h_n$ are the spherical Hankel functions,
and $\phi_{k,nm}$ are the coefficients of $f_k(\vec{\psi})$ projected on the spherical harmonic basis,
\begin{equation}\label{eq:coeff}
    \phi_{k,nm} = \int_{\mathbb{S}^2} f_k(\vec{\psi})Y_n^m(\vec{\psi})\txt{d}\vec{\psi}.
\end{equation}
Equation~\eq{decode} is the sound pressure of frequency band $k$. In the time domain,
the received sound is a summation over all frequency bands, namely,
$s(r,\vec{\psi},t) = \sum_k p_k(r,\vec{\psi})e^{-i\omega_k t}$, where $\omega_k$ is the frequency
corresponding to the wave number $k$.
Correspondingly, $\phi_{k,nm}$ in the frequency domain can
be rewritten in the time domain using the Fourier transform,
\begin{equation}\label{eq:coeff_time}
    \phi_{nm}(t) = \sum_k \phi_{k,nm}e^{-i\omega_k t} = \int_{\mathbb{S}^2} f(\vec{\psi},t)Y_n^m(\vec{\psi})\txt{d}\vec{\psi}.
\end{equation}
where $f(\vec{\psi},t)$ is the directional distribution of sound source signals in time domain.

In essence, ambisonic audio records the coefficients
$\phi_{nm}(t)$ (normalized by a constant) up to a certain order $n$.
At runtime, an ambisonic decoder generates audio signals output to
speaker channels (such as stereo and 5.1) according to~\eq{decode}
together with a head-related transfer function model.
Currently, all the mainstream \tsd video players, such as Youtube and Facebook video players,
support only first order ambisonics, which takes four channels of signals corresponding to
$\phi_{nm}$ at $n=0,m=0$ and $n=0,m=-1,0,1$.

\paragraph{Generating ambisonic channels}
Let $s_i(t)$ denote the dry audio signals.  Using the ambisonic model, we view
each early ray as a directional sound source, whose signal $s(t)$ is the dry audio
convolved with its ERIR component (i.e., $s(t)=s_i(t)*H_{r,\vec{\theta}}(t)$, where $H_{r,\vec{\theta}}(t)$
is introduced in \eq{ht}). Because this ray
comes from direction $\vec{\theta}$, we model the corresponding $f(\vec{\psi},t)$
in~\eq{coeff_time} as a Dirac delta distribution scaled by its incoming signal $s(t)$:
$f(\vec{\psi},t) = \delta(\vec{\psi}-\vec{\theta})s(t)$. Then, the audio data
due to early reverberation at each ambisonic channel is
\begin{equation}\label{eq:coeff2}
    \phi_{nm} = \int_{\mathbb{S}^2} \delta(\vec{\psi}-\vec{\theta})s(t)Y_n^m(\vec{\psi})\txt{d}\vec{\psi}
    = Y_n^m(\vec{\theta})\left(s_i(t)*H_{r,\vec{\theta}}(t)\right).
\end{equation}
In our examples, we compute $\phi_{nm}$ only up to the first order because of the
limitation in current \tsd video players. This results in four channels of
signals (named as the $W$-, $X$-, $Y$-, and $Z$-channel), and their corresponding
$Y_n^m$ are $\frac{1}{\sqrt{2}}$, $\cos\theta\cos\phi$, $\sin\theta\cos\phi$,
and $\sin\phi$ respectively, where $\theta$ and $\phi$ are the azimuthal and zenith angle of
the direction $\vec{\theta}$.
We iterate through all incoming rays collected in~\secref{direction_aware}, compute
their $\phi_{nm}$ using~\eq{coeff2} and accumulate them into corresponding channels.

Meanwhile, the LRIR component produces audio signals $s_\txt{L}(t)=s_i(t)*H_\txt{L}(t)$.
We model $s_\txt{L}(t)$ as sound signals coming uniformly from all directions according to our
observation of energy isotropy in LR (recall~\secref{erDuration}). Then, $f(\vec{\psi},t)$ in~\eq{coeff_time}
becomes a direction independent function,
$\frac{1}{4\pi}s_\txt{L}(t)$. In this case, the $W$-channel is accumulated by $\frac{1}{\sqrt{2}}s_\txt{L}(t)$,
while the $X$-, $Y$-, and $Z$-channels are not affected.

After this step, the four channels of audio data are encapsulated into the \tsd
video. Our method can readily produce ambisonic audio with
higher-order channels for future \tsd video players.

\begin{figure}[t]
    \includegraphics[width=0.99\hsize]{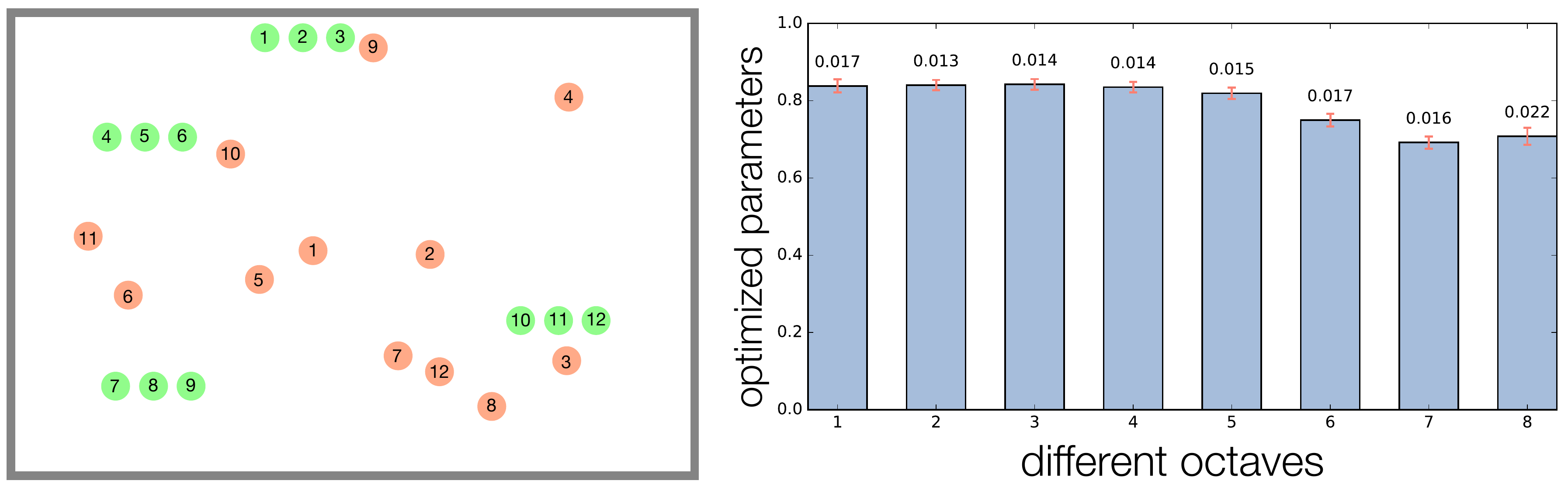}
    \vspace{-3mm}
    \caption{{\bf Position independence.} 
    We recorded 12 IRs in a room at different source and receiver locations,
    and perform our material estimation (\secref{mat_ana}) separately using each of the IRs.
    (left) We visualize the source (in green) and listening (in orange) positions 
    used in each IR measurement indicated by the numbers inside the dots.
    (right) For each measurement, we optimize the material parameters, and plot the average
    value in each octave band (x-axis), along with error bars (indicating one
    standard deviation) shown on top
    of the bars. This plot shows that the material estimation is virtually independent from the choice
    of source and receiver locations.
    \label{fig:position_independence}}
    \vspace{-3mm}
\end{figure}

\section{Results}
\label{sec:results}
We now  validate our method, and present several useful
applications. To fully appreciate our results, we encourage readers to watch our
accompanying video. Our results were computed on a 4-core Intel i7 CPU. 
Our system, including acoustic simulation, material optimization, determination of $T_{ER}$, frequency modulation, and ambisonic encoding, takes $\approx$ 10-20 seconds.
In addition to our main supplemental video, we also provide our raw 360 videos and instructions
in a supplemental zip file for full immersive experience.
\camera
{
We use Ricoh Theta V with built-in first-order ambisonic audio microphones for
the recordings.
}

\subsection{Validation}
\label{sec:validation}

\paragraph{Directionality of LRIR} 
While the common assumption that the LRIR is diffuse spatially 
has been exploited in previous methods like Raghuvanshi and Snyder~\shortcite{Raghuvanshi:2014:PWF},
its isotropy with respect to direction has received less attention.
We  provide evidence through room acoustic measurement using 
a highly directional ``shotgun'' microphone. The details are described in
\figref{LRIR_direc}. 
An additional plot that also appears in the supplemental video is explained 
in \figref{uniform_srir}.



\paragraph{Robustness of material parameter estimation}
Part of the ease of our method rests on the fact that we only need one recorded
impulse response per room using a conventional mono microphone, 
and that the positions of the source and receiver when recording do not matter.
Figure~\ref{fig:position_independence} demonstrates the negligible impact these
positions have on our IR measurement and material estimation steps.
The same experiment also confirms that the LRIRs in all the measured IRs 
closely match each other. 
This bolsters the common assumption that the LRIR is spatially diffuse.

\begin{figure}[t]
    \includegraphics[width=0.95\hsize]{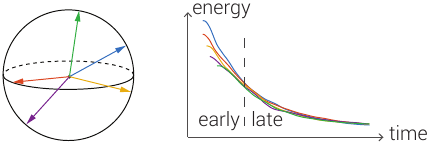}
    \caption{\textbf{Directional energy response.} We measure the directional energy response $h_{\vec{\theta}}(t)$
    along five incoming directions (left) using a directional shotgun microphone. These measured $h_{\vec{\theta}}(t)$ (right)
    have different ER parts, but as time increases their LR tails converge.
    \label{fig:uniform_srir}}
    \vspace{-5mm}
\end{figure}


\begin{figure}[b]
\vspace{-2mm}
    \includegraphics[width=0.8\hsize]{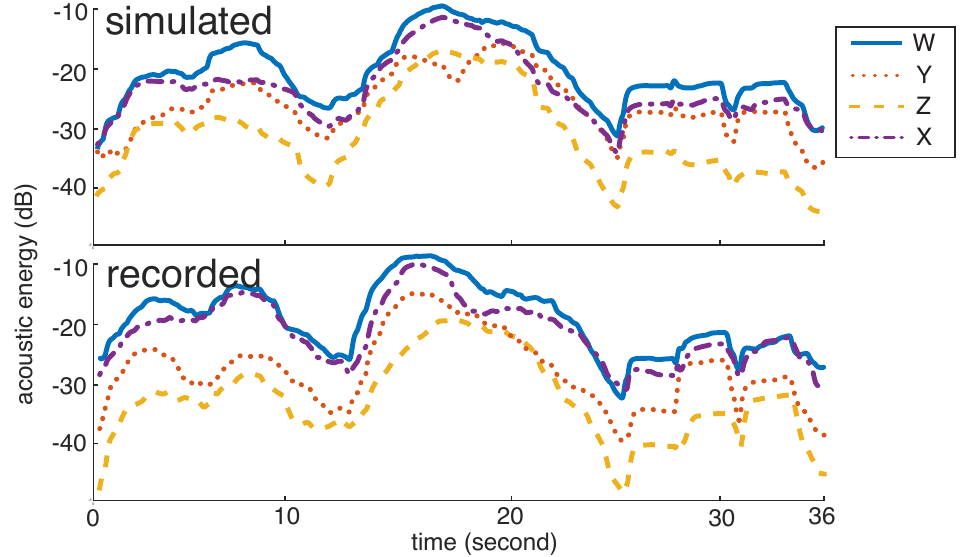}
\vspace{-3mm}
\caption{\camera{\textbf{Coherent temporal variation.}  
Here we show the time-varying acoustic energy levels of simulated (top)
and recorded (bottom) audio. The four curves correspond to the four ambisonic
channels.
\label{fig:matchRecordings}}}
\end{figure}

\begin{figure}[t]
    \includegraphics[width=\hsize]{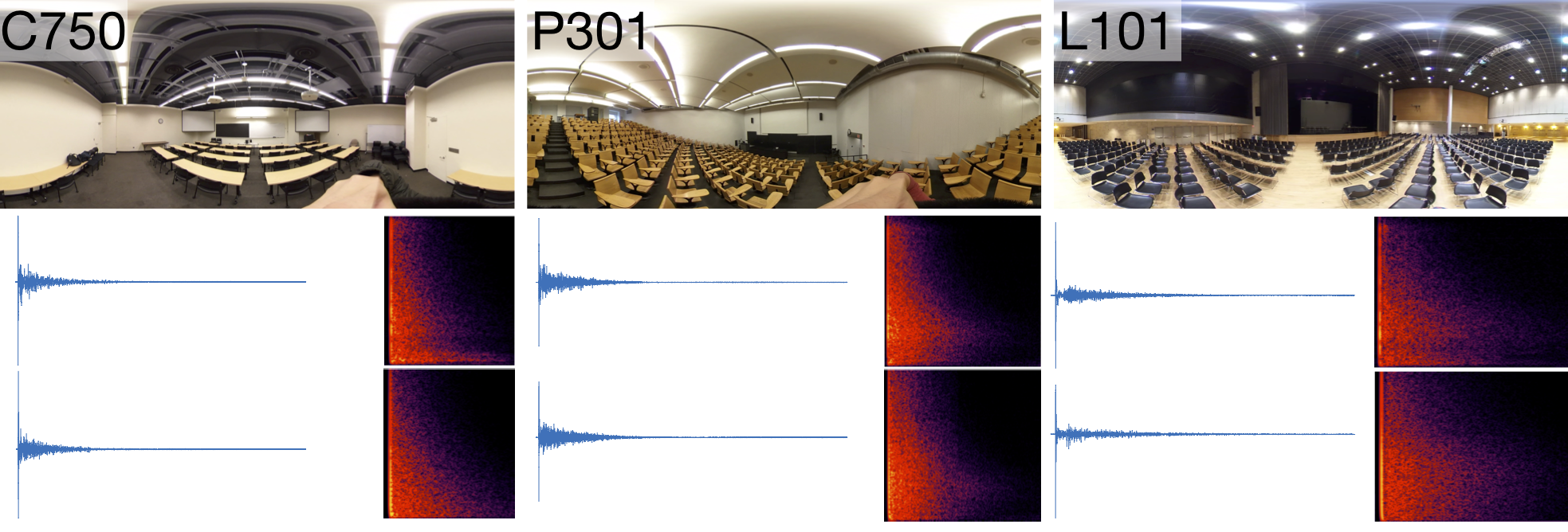}
    \caption{{\bf Matching recorded IRs.} Our method (bottom) produces IRs that
    match closely recorded IRs (middle) for three different cases (top).
    Shown here are IRs of three distinct rooms. Their sizes, shapes, and
    materials vary largely (see Table~\ref{tab:stats}).
    We refer to the supplemental materials that include audio clips of these IRs.
}\label{fig:singleroom-match}
\vspace{-4mm}
\end{figure}

\paragraph{Agreement with recordings}
We demonstrate that our algorithm can faithfully match recorded audio using ambisonic microphones.
We compare recorded audio to the \tsd audio synthesized by our method.
In several rooms of varying size, our results match very well with the
recordings (Figure~\ref{fig:singleroom-match}). Again, please see our
accompanying video to appreciate the high level of agreement our method has with
recordings. To highlight the match with recordings, we stitch
the recorded and synthesized audio side-by-side, to show the nearly seamless transitions.

\camera{
\paragraph{Temporal Coherence}
Figure~\ref{fig:matchRecordings} shows the temporal variation of
the acoustic energy of the four ambisonic channels 
on our synthesized and recorded audios.
In this validation, the camera first moves towards a sound source and
then moves away. 
The simulated and recorded audio exhibit similar variations.
}

\subsection{Applications}\label{sec:app}
Our approach enables several novel applications which make spatial audio for \tsd videos
easier to work with.

\begin{figure}[t]
    \includegraphics[width=\hsize]{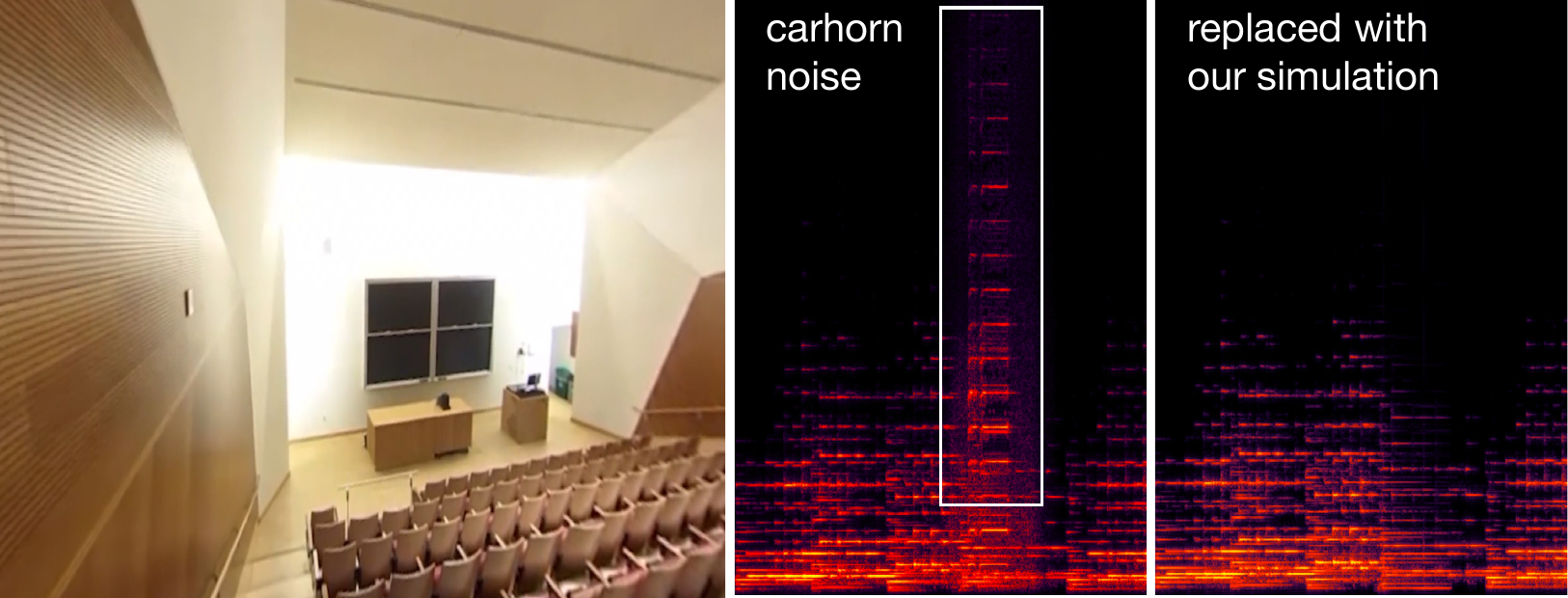}
    \caption{{\bf Audio replacement.} While recording sound in a classroom, there
    was an unwanted car horn outside. The car horn overlapped in frequency with
    our desired audio, which makes removing it challenging. Using our method to
    resimulate the dry audio provides noise free audio that sounds as if it was
    recorded in the scene.}
    \label{fig:audioreplacement}
\end{figure}

\paragraph{Audio replacement in \tsd video}
While ambisonic microphones can be used to record spatial audio directly, 
they have limited use in the production pipeline. Many sounds are added to
videos in post production, instead of during the video shooting.
Our method allows adding sound to \tsd video during post-production in a
realistic spatialized fashion. We have done this in various classrooms, lecture
halls, and auditoriums with varying sizes and reverberation characteristics,
some of which are shown in \figref{singleroom-match}.
An additional, concrete application is the removal of unwanted sound, shown in \figref{audioreplacement}.
During one of our recordings, an unwanted car horn came from outside. Noise
removal can be challenging, especially for non-stationary sources that overlap
in frequency. Our method allows resimulating the desired dry audio, making it
sound as if it was recorded in the same room, but with no noise.

\begin{table}[b]
    \centering
    \resizebox{1.\linewidth}{!} {
    \begin{tabular}{l|rrrr}
    \whline{1.0pt}
                        &  size (m)    &  $T_\txt{ER}$ (msec) & \camera{\# planes} & type    \\
    \whline{0.8pt}
    \textsc{CEPSR401}       & 15$\times$20$\times$6   & 44  & \camera{ 6 }& indoor           \\ \rowcolor[gray]{0.93}
    \textsc{CEPSR620}        & 4$\times$6$\times$3    & 21 & \camera{11}  & indoor          \\
    \textsc{CEPSR750}      & 11$\times$8$\times$4     & 37  & \camera{6} &   indoor       \\\rowcolor[gray]{0.93}
    \textsc{Pupin301}     & 12$\times$17$\times$7     & 68 & \camera{6} & indoor         \\
    \textsc{NWC501}   & 11$\times$15$\times$6       & 46  & \camera{6} & indoor      \\ \rowcolor[gray]{0.93}
    \textsc{Lerner101}      & 40$\times$60$\times$12   & 121 & \camera{6} & indoor      \\
    \textsc{Hallway}      & 2$\times$15$\times$5    & 40 &\camera{ 17} &  multiroom       \\\rowcolor[gray]{0.93}
    \textsc{Outdoor}   &  70$\times$50 & 164 &\camera{ 21} & outdoor          \\
    \whline{1.0pt}
    \end{tabular}
    }
    \caption{{\bf Example Statistics.} \camera{Here we list the estimated
        dimension, early cutoff time $T_\txt{ER}$, number of planes used in our
    simulation, and type of scene.}
    }
    \label{tab:stats}%
\end{table}

\begin{figure}[t]
    \vspace{-0.5mm}
    \includegraphics[width=0.99\hsize]{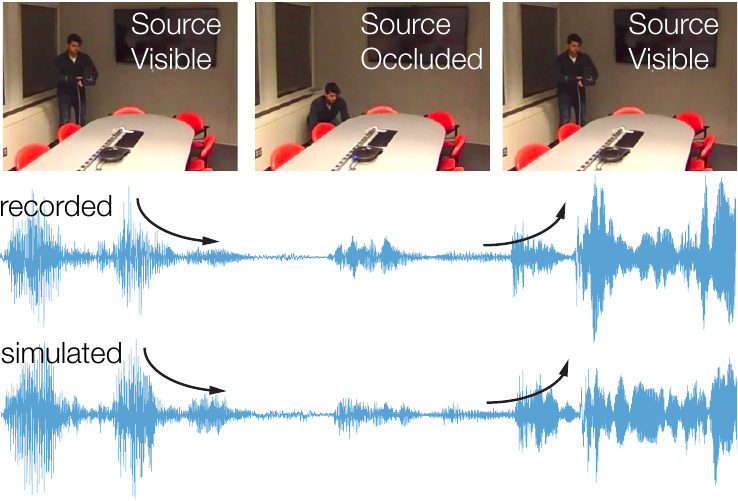}
    \vspace{-3mm}
    \caption{{\bf Geometric effects: occlusion.} As a sound source moves below a table, it
    exhibits a low-pass muffling effect due to direct sound being blocked. Our
    method captures this effect. }
    \label{fig:desk}
    \vspace{-3mm}
\end{figure}

\paragraph{Geometric effects} One of the main benefits our method provides to
\tsd video editors is the ability to automatically capture geometric effects.
This can easily be seen when geometry occludes the source or receiver. 
In this example, we moved a speaker above and below a table, causing the sound to
become muffled. Our method captures this effect automatically
(Figure~\ref{fig:desk}). Instead of painstakingly adjusting amplitude and frequency to
approximate shadowing, sound editors can now just apply a geometric filter.

\begin{figure}[b]
    \vspace{-2mm}
    \begin{overpic}[width=0.99\hsize]{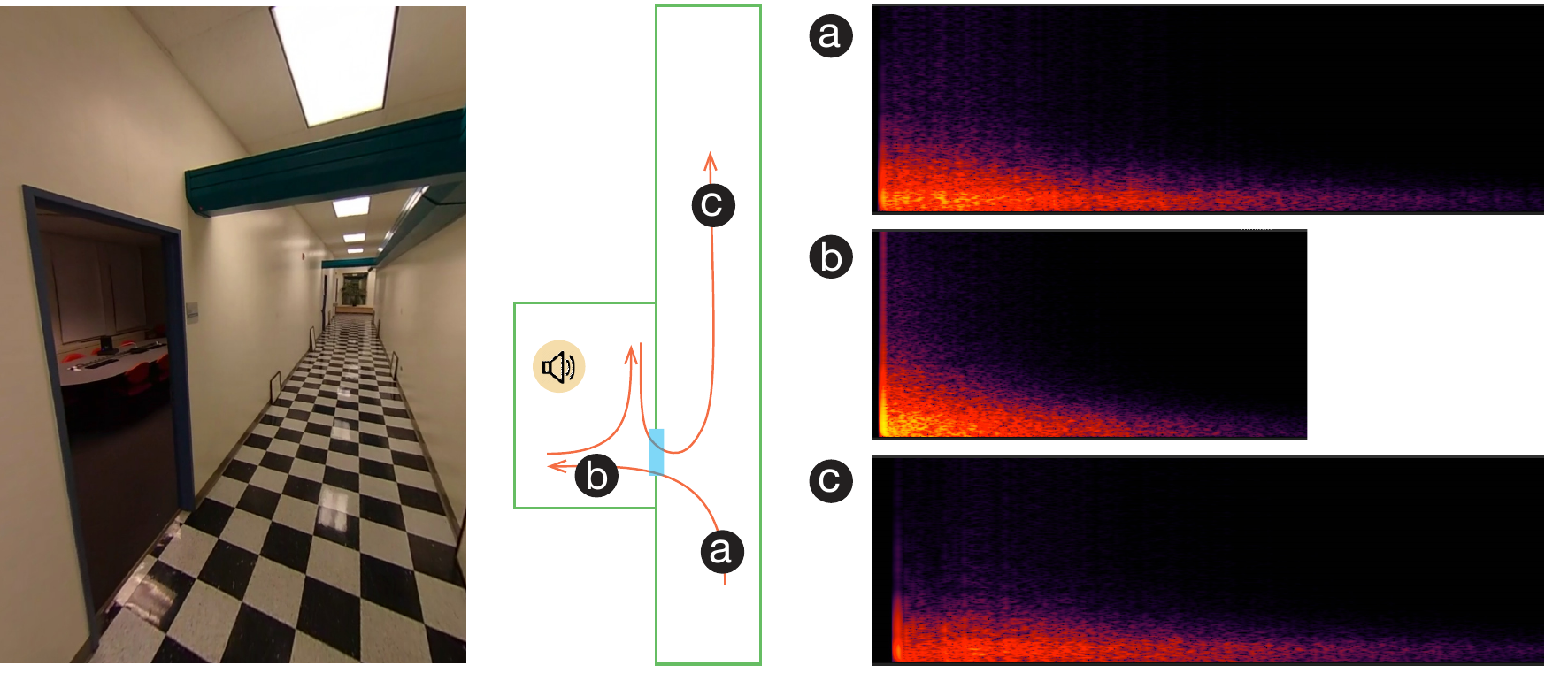}
    \put(13.5,-2.5){(a)}
    \put(40,-2.5){(b)}
    \put(78,-2.5){(c)}
    \end{overpic}
    \vspace{-1mm}
    \caption{{\bf Connected rooms.} As a listener moves between rooms, the
    reverberation changes, and strong geometric shadowing effects are heard. Our
    method naturally works in these cases, requiring only one IR measurement in
    each room.
    (a) A photograph of the multi-room scene. (b) The layout of the rooms.
    (c) The spectrograms of the synthesized IRs at three distinct locations.
    } \label{fig:multiroom}
\end{figure}

\paragraph{Extension to cross-room propagation}
An even stronger geometric effect happens when a source or listener moves
between rooms. This can cause very different sound due to the small opening
between rooms, and different reverberation in each room. A simple extension of
our method to two rooms is demonstrated (\figref{multiroom}).
Consider a source $s$ located in room 1 and a listening location $d$ in room 2.
Just like single rooms, 
the (directional) ERIR $H^{12}_E$ between $s$ and $d$ is computed from simulated rays with spatial effects.
For the LRIR, we recorded an IR once in each room, $H^1$ and $H^2$.
We then compute the propagated IR between two rooms as 
\begin{equation}\label{eq:2IR}
    H^{12}_L = a\sum_{p\in A}\left(H^1_{E,\txt{s}\to\txt{p}}*H^2_L + H^2_{E,\txt{d}\to\txt{p}}*H^1_L + H^1_L*H^2_L\right),
\end{equation}
where $p$ are locations uniformly sampled in the planar region of the door $A$,
$a$ is the effective area of each sampled location $p$,
$H^1_L$ and $H^2_L$ are the LR components of the recorded IRs in each room,
and $H^1_{E,\txt{s}\to\txt{p}}$ and $H^2_{E,\txt{d}\to\txt{p}}$ are simulated
ERIRs from $s$ to $p$ in room 1 and from $d$ to $p$ in room 2.
The derivation of~\eq{2IR} is presented in Appendix~\ref{app:2IR}.
This formulation is similar to~\cite{Stavrakis:2008:Graphs}.
Therefore, our algorithm could be easily extended to a general graph of connected
rooms using their algorithm.



\begin{figure}[t]
    \vspace{-2mm}
    \includegraphics[width=0.99\hsize]{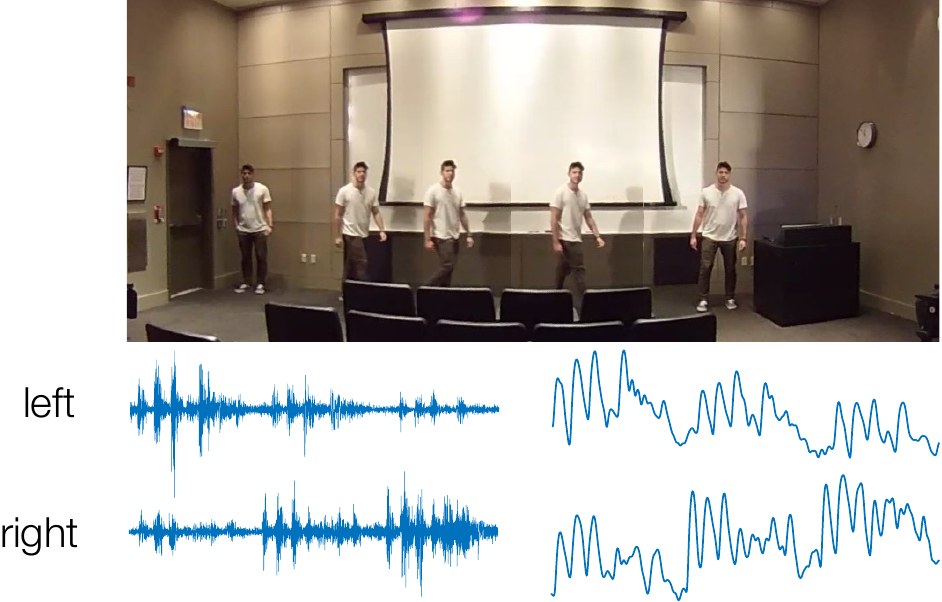}
    \vspace{-2mm}
    \caption{{\bf Re-spatialization.} From recorded mono audio of a person
    talking while moving around a room, we can re-spatialize the sound. By using
    our method to compute the energy distribution due to the moving source, we
    distribute the mono sound energy appropriately. (top) Sound source moving
    from left to right in the camera frame. (bottom) The sound waveform (left)
    and the energy (right) after binauralizing our spatialization. Notice how
    the sound follows the source, moving from left to right.}
    \label{fig:davis-mono2ambi}
    \vspace{-3mm}
\end{figure}

\paragraph{Re-spatialization of mono audio}
The final application we present is a way to apply spatial effects to
\emph{in-situ} recorded mono audio, i.e., audio recorded in a room with reverberation. 
This problem is similar in spirit to the conversion of a 2D film into a 3D film without 
refilming it --- a popular problem in the film industry.
Theoretically, re-spatializing the audio could be done by \emph{deconvolving}
the impulse response from the recorded audio, to obtain the original (``dry'') source
audio. The dry audio could then be spatialized with our method. However,
deconvolution is a very ill-conditioned process and is difficult in practice.
Instead, we present an ad-hoc effect that can give some spatial impression.
Given a room model with estimated materials, we perform a full IR simulation and
store the propagated rays. We can then take the input mono-channel audio and distribute its
energy over the sphere to match the energy of the computed rays. While not
fully principled, it provides a plausible effect and works well
in many cases, shown in \figref{teaser} and \figref{davis-mono2ambi}.

\section{Conclusion}

We have presented a method for adding realistic, scene-aware spatial audio to \tsd videos. By
combining simulated early reflections with recorded late reverberation, our
method is extremely fast and matches recorded audio well. It provides a
practical way to incorporate geometric effects during audio post-production,
requiring only a standard mono microphone and a \tsd camera. We believe this
will enable the next generation of sound design for emerging immersive content.

\paragraph{Limitations and future work}
A major limitation to proper viewing of spatial audio currently is the lack
of personalized head-related transfer functions. These functions describe how
our head and ear geometry modifies sound before it reaches our ear drums, which
is how humans detect directionality of sound. These functions are unique to
individuals, but are laborious to measure. While the common/average models that
current \tsd video players use give a spatial impression, we expect 
the accuracy continue to
increase in the future as personalized HRTFs become easier to
obtain.
\camera{
We note again that our method supports an arbitrary order of ambisonics, while
most current players support only first order.}

\begin{figure}[t]
    \vspace{-2mm}
    \includegraphics[width=0.99\hsize]{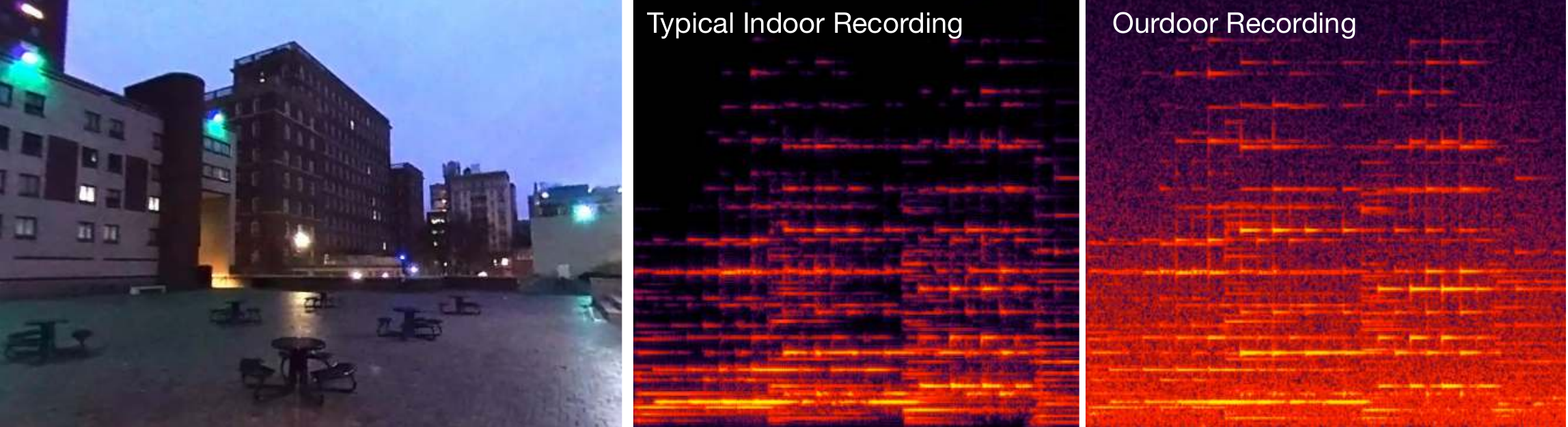}
    \vspace{-3mm}
    \caption{{\bf Challenging outdoor case.} 
    We applied our method to outdoor \tsd videos. The major challenge is
    recording noise, due to e.g., environment and wind, which makes an exact match
    of our synthesized audio to the ambisonic recordings very challenging.
    However, the results sound plausible (see video).
    (left) An outdoor \tsd recording scene at 6AM in the morning. (right) The recorded audio severely contaminated by noise.
    (middle) A typical indoor recording with much less noise.
    }
    \label{fig:outdoor}
    \vspace{-3mm}
\end{figure}

Our method requires a good impulse response to work well. While much
easier and faster than directly measuring acoustic properties of scene
materials, it is still an extra step that requires access to the original room
where the video was recorded. Future work could examine inferring an impulse
response from the audio in the video.
Large spaces such as outdoor scenes are challenging. 
The large amount of uncontrollable noise makes it
difficult for our method to match recordings exactly, as shown in
\figref{outdoor}. However, this could also be seen as a strength of our method:
the ability to re-simulate only the audio sources of interest, noise free. 

\camera{
To ease the IR measurement,
only one measurement per room is needed in our method.  
However, this limits our ability to detect material differences among
different indoor regions. While a single measurement appears to be 
sufficient in our method, 
more precise estimation of wall materials may be necessary
in order to simulate an impulse response accurately. 
This could possibly be achieved
using multiple recordings at different locations.
 }

  Currently we only model the major walls and obstacles in the scene, ignoring
  most other objects like chairs. While it is reasonable to drop small features
  when the sound wave length is large enough, we indeed oversimplify the
  reconstructed geometry. One issue occurs when the listening location becomes
  too close to an object that we do not model/optimize. In this case, the
  synthesized audio sounds may characteristically differ from the recordings. In
  our experiments, we found that keeping a safe distance between unmodelled
  objects prevents this discrepancy.
In the future, we wish to investigate
the impact of accurate geometric modeling on the optimization process as well as the
resulting audio.

\camera{
Currently, we separate ER and LR parts based on the point where the sound field 
becomes directionally diffuse.  
It remains an open question as to what the optimal way is to 
separate the IR, 
since the separation time depends on many factors,
 such as acoustic energy, directional distribution, number of sound sources, and others. 
}

Realistic spatial audio authoring in \tsd videos is an exciting and challenging
research field. Thanks to recent hardware developments and surging interest in
virtual reality, we expect to see an increased
demand for immersive \tsd audio.
Our scene-aware audio is a first step towards the practical application of a more immersive audio-visual experience.
In order to further advance the audio quality,
still more accurate and efficient methods are required.
\camera{
Provided the current active research toward realtime GA simulation,
an interesting future work is to extend our system with realtime simulation 
for virtual and augmented reality applications.}


We believe that an intuitive spatial audio editing pipeline 
will go a long way to advance virtual reality audio editing. 
Unlike mono-channel or stereo audio, high-order
ambisonics have quadratically increasing number of channels, i.e., 1st order has
4, 2nd order has 9, and so on. While low-level mixing and stitching works on one
or two channels for traditional audio, we argue that a higher-level abstraction
of the audio editing process can help users access the full potential of spatial audio.
Our work abstracts the manipulation of different channels to intuitive
concepts such as the virtual sound source location and listening location,
allowing designers to think more about the scene and less about waveform
editing.
Lastly, we look forward to other avenues where spatial audio will enhance the user experience.







\begin{acks}

We thank Chunxiao Cao for discussing and sharing his bidirectional sound simulation code, 
Carl Schissler for sharing the ``infinite'' audio file,
James Traer for discussion on IR measurement, 
and Henrique Maia for proofreading and voiceover.
This work was supported in part by the National Science Foundation (CAREER-1453101), 
SoftBank Group, and generous gift from Adobe. Dingzeyu Li was 
partially supported by an Adobe Research Fellowship.
\end{acks}

\bibliographystyle{ACM-Reference-Format}
\bibliography{paper}

\appendix
\section{Derivation for Cross-Room IR}\label{app:2IR}
Consider a source $s$ located in room 1 and a listening location $d$ in room 2.
The IR between $s$ and $d$ is the result of propagating sound though the door, 
and thus can be written as
\begin{equation}\label{eq:IRconv}
H^{12}_{\txt{s}\to\txt{d}}(t) = \int_S
H^1_{\txt{s}\to\txt{p}}(t)*H^2_{\txt{p}\to\txt{d}}(t) \, \txt{d}S(p),
\end{equation}
where $S$ is the door area that connects two rooms (the semi-transparent blue region
in \figref{multiroom}-b), and $p$ is a point located in the door region.
$H^1_{\txt{s}\to\txt{p}}(t)$ and $H^2_{\txt{p}\to\txt{d}}(t)$ are the IRs between
$s$ and $p$ in room 1 and between $p$ and $d$ in room 2, respectively. They can be approximated
as concatenations of the simulated ERIR and measured LRIR in each room, namely,
\begin{equation}
    H^1_{\txt{s}\to\txt{p}} = H^1_{E,\txt{s}\to\txt{p}} + H_L^1\;\textrm{and}\;
    H^2_{\txt{p}\to\txt{d}} = H^2_{E,\txt{d}\to\txt{p}} + H_L^2,
\end{equation}
where $H_L^1$ and $H_L^2$ are the LR components of the IRs recorded in each
room independently (following \secref{audioAnalysis}), and $H^1_{E,\txt{s}\to\txt{p}}$
and $H^2_{E,\txt{d}\to\txt{p}}$ are simulated ERIRs between $s$ and $p$
in room 1 and between $d$ and $p$ in room 2. We note that here we use 
$H^2_{E,\txt{d}\to\txt{p}}$ but not $H^2_{E,\txt{p}\to\txt{d}}$ because they are 
the same due to acoustic reciprocity. Then, the integrand in~\eq{IRconv} becomes
\begin{equation}
    \underbrace{H^1_{E,\txt{s}\to\txt{p}}*H^2_{E,\txt{d}\to\txt{p}}}_\text{ERIR} + 
    \underbrace{H^1_{E,\txt{s}\to\txt{p}}*H^2_L + H^2_{E,\txt{d}\to\txt{p}}*H^1_L + H^1_L*H^2_L}_\txt{LRIR},
\end{equation}
where the ERIR is replaced by with our acoustic simulation.
After we discretize the door region using sampled points, the LRIR becomes 
the expression~\eq{2IR}.

\end{document}